\documentclass[11pt]{article}

\usepackage[margin=1in]{geometry}
\usepackage{amsmath}
\usepackage{amssymb}
\usepackage{booktabs}
\usepackage{graphicx}
\usepackage{subcaption}
\usepackage{array}
\usepackage{tabularx}
\usepackage[numbers,sort&compress]{natbib}
\usepackage[colorlinks=true,allcolors=blue]{hyperref}
\usepackage[capitalise,nameinlink]{cleveref}

\crefname{figure}{Fig.}{Figs.}
\Crefname{figure}{Figure}{Figures}
\crefname{table}{Table}{Tables}
\Crefname{table}{Table}{Tables}
\crefname{equation}{Eq.}{Eqs.}
\Crefname{equation}{Equation}{Equations}
\crefname{subfigure}{Fig.}{Figs.}
\Crefname{subfigure}{Figure}{Figures}

\title{Interfacial Roughness Spectra and Finite-Depth Salt-Finger Mixing at a Two-Layer Thermohaline Interface}
\author{Sriram P. Kalathoor\\
Daniel Guggenheim School of Aerospace Engineering\\
Georgia Institute of Technology\\
\texttt{sriram2@gatech.edu}}
\date{}

\begin{document}

\maketitle

\begin{abstract}
Salt fingering drives diapycnal scalar exchange across thermohaline interfaces that are statically stable but double-diffusively unstable. Oceanic interfaces are finite-depth structures and may carry roughness inherited from waves, shear, intrusions, or prior mixing. We test how the horizontal spectrum of that roughness controls the route from a two-layer interface to a finite-depth salt-finger plume forest. Direct simulations of the modeled Boussinesq equations are performed at \(\mathrm{Pr}=7\), \(\tau=0.01\), and \(\mathrm{R}_\rho=1.2\), with matched domain, grid, amplitude, boundary treatment, and analysis measures. The imposed spectra are high-annulus, low-mode, and mixed; a second mixed realization tests robustness.

The imposed spectrum selects distinct routes to vertical exchange. High-annulus roughness remains compact and branch-locked through \(t=60\), without a tracked broad-branch transition. Low-mode roughness begins on the broad branch, produces the strongest salinity transport at \(t=45\), and reaches the finite-depth boundary region first. Mixed roughness follows a velocity-led pathway: vertical velocity selects the broad branch before salinity, while salinity develops the richest planform spectral population. At \(t=45\), the mixed salinity effective mode count is \(86.66\), compared with \(3.26\) for high-annulus forcing and \(5.46\) for low-mode forcing. Angular and signed-branch measures show branch-dependent diagonal organization, and probe/volume measures show that local plume-passage asymmetry does not imply large global upper/lower imbalance. The replicate preserves the mixed route with shifted transition times. Thus a finite-depth thermohaline interface can retain spectral memory, controlling whether salt-finger mixing remains localized, penetrates rapidly, or forms a scalar-rich plume forest through delayed modal handoff.

\end{abstract}

\section{Introduction}
\label{sec:introduction}

Salt fingering is the double-diffusive instability that occurs when a stabilizing temperature stratification and a destabilizing salinity stratification coexist in a statically stable density field. Because heat diffuses faster than salt, a displaced warm salty parcel can lose heat more rapidly than salt, become denser than its surroundings, and continue moving away from its original level. This salt-fountain picture was introduced in the early theoretical literature \citep{stommel1956perpetual,stern1960salt}. Laboratory and theoretical studies then placed the mechanism into the broader framework of convection driven by combined heat and salt gradients
\citep{turner1964newcase,turner1973buoyancy,turner1974double,turner1985multicomponent}. In modern notation, the local
fingering condition is commonly written in terms of the density ratio \(\mathrm{R}_\rho\) and diffusivity ratio \(\tau\): salt fingering occurs when \(1<\mathrm{R}_\rho<1/\tau\) \citep{schmitt1979growth,radko2013double}.

The oceanic importance of that local criterion comes from the fact that temperature and salinity gradients occur in structured water masses, not only in idealized uniform backgrounds. Turner-angle analyses and water-mass analyses show where double diffusion is favored in the ocean
\citep{ruddick1983practical,you2002turner}. Thermohaline relationships in the
central water and tracer-release measurements connect salt fingering to measurable diapycnal exchange \citep{schmitt1981form,schmitt2005enhanced}. Reviews emphasize that double diffusion can alter both local scalar fluxes and larger-scale water-column structure \citep{schmitt1994double,gargett2003differential}. The physical question therefore extends beyond whether fingers grow. Their nonlinear organization changes interfacial geometry, active-layer thickness, and vertical exchange over the available depth.

Layered double-diffusive systems add a second piece of motivation. Oceanic and geophysical double-diffusive structure can appear as intrusions, steps, staircases, and finite-thickness layers. The diffusive and fingering regimes have different scalar orderings, but both show that double-diffusive dynamics can couple local transport to layer-scale structure
\citep{huppert1981double,kelley2003diffusive}. Arctic and marginal-sea
staircase observations demonstrate that persistent thermohaline layering is a real oceanic state rather than only a mathematical idealization
\citep{timmermans2008ice,shibley2017arctic,durante2019permanent}. The present
work does not claim staircase formation. Instead, those studies motivate the need to separate local plume growth, active-layer thickening, and true layer formation when analyzing finite-depth simulations.

Canonical salt-finger theory provides the mechanism against which the present finite-depth problem must be checked. Collective-instability theory shows that fingering can drive larger-scale motions rather than remaining a passive array of independent columns \citep{stern1969collective}. Salt fingers and convecting layers were connected early in the theory of double-diffusive interfaces \citep{stern1969layers}. Finite-length and secondary-instability analyses further show that primary fingers need not remain long, steady, and laminar \citep{holyer1984stability,kunze1987limits}. These results justify diagnosing nonlinear route selection through modal organization, vertical reach, and scalar transport rather than through instantaneous finger width alone.

Direct simulations and mean-field studies provide the closest numerical background. Three-dimensional salt-finger simulations have resolved the transition from primary fingers through secondary instability to chaotic convection \citep{simeonov2009dns}. Homogeneous-gradient fingering studies have quantified small-scale fluxes and large-scale instabilities
\citep{traxler2011dynamics}. Companion simulations and mean-field theory show
how fingering turbulence can reorganize into thermohaline staircases
\citep{radko2003mechanism,stellmach2011dynamics,radko2014recipes}. Related numerical work in
low-Prandtl and astrophysical regimes demonstrates that double-diffusive transport and layer formation remain sensitive to parameter regime and scalar diffusivity ordering \citep{brown2013chemical,garaud2018double}.

These studies establish the checks needed here: the modeled state must lie in the fingering regime, the resolved flow must move beyond laminar columns, transport must be measured directly, and scale reorganization must be distinguished from staircase formation or boundary interaction. They also leave a finite-depth gap. Most canonical DNS and flux-law studies begin from homogeneous gradients or idealized mean states, whereas many oceanic and laboratory interfaces enter the fingering regime with finite-amplitude roughness inherited from waves, shear, intrusions, remnant turbulence, or earlier deformation. The initial roughness spectrum can therefore select the route followed by the plume forest and the way the interface exchanges salt with adjacent layers.

The present study tests that route-selection idea with matched direct simulations of the modeled Boussinesq equations. We initialize a two-layer thermohaline interface with controlled finite-amplitude roughness spectra motivated by interfaces whose shape is inherited from waves, shear, intrusions, remnant turbulence, or prior interfacial deformation. The simulations differ only in the imposed interfacial spectrum and, for one robustness calculation, in the phase and noise realization. The three primary spectra form a compact controlled family: a low-mode spectrum, a high-annulus spectrum, and a mixed spectrum formed from both.

The comparison shows that the roughness spectrum controls the nonlinear route. High-annulus forcing remains compact and branch-locked. Low-mode forcing selects a broad branch almost immediately, gives the largest interior-comparison transport, and reaches the finite-depth boundary region first. Mixed forcing follows a distinct route in which vertical velocity selects the broad branch before salinity while the scalar field develops the richest planform spectral population. A second mixed-spectrum realization preserves this route with shifted transition timing.

The contribution is a physical-oceanography result about finite-depth route selection. The imposed interfacial spectrum controls morphology, modal handoff, scalar spectral population, transport, active-layer thickening, local plume-passage asymmetry, and boundary approach. Separating the interior-growth window from later boundary approach prevents route selection from being confused with relaxation-zone interaction.

\section{Problem Setup and Numerical Approach}
\label{sec:problem_setup}

\subsection{Modeled equations and nondimensional parameters}
\label{sec:setup_equations}

The simulations solve the incompressible Boussinesq equations with active temperature and salinity scalars. In the nondimensional form used here,
\begin{align}
  \nabla \cdot \boldsymbol{u} &= 0, \\
  \partial_t \boldsymbol{u}
    + \boldsymbol{u}\cdot\nabla\boldsymbol{u}
    &= -\nabla p + \mathrm{Pr} \nabla^2 \boldsymbol{u}
       + \mathrm{Pr}\, b\,\boldsymbol{e}_z, \\
  \partial_t T + \boldsymbol{u}\cdot\nabla T
    &= \nabla^2 T, \\
  \partial_t S + \boldsymbol{u}\cdot\nabla S
    &= \tau \nabla^2 S,
\end{align}
where \(\boldsymbol{u}=(u,v,w)\), \(z\) is vertical, \(\mathrm{Pr}\) is the Prandtl number, and \(\tau\) is the salinity-to-thermal diffusivity ratio. The temperature diffusivity is the reference diffusivity, so \(\mathrm{Pr}\) sets the viscosity and \(\tau\) sets the salinity diffusivity. The linear equation of state is written with unit thermal expansion and haline contraction in nondimensional variables, so that
\begin{equation}
  b = T - S .
\end{equation}
The control parameters are
\begin{equation}
  \mathrm{Pr}=7, \qquad \tau=0.01, \qquad \mathrm{R}_\rho = 1.2 .
\end{equation}
With the present normalization, the temperature jump is \(\Delta T=1\) and the salinity jump is \(\Delta S=\Delta T/\mathrm{R}_\rho\). Thus \(1 < \mathrm{R}_\rho < 1/\tau\), placing the interface in the salt-fingering regime. All coordinates, times, scalar amplitudes, velocities, fluxes, wavelengths, and envelope widths reported below are nondimensional unless explicitly stated otherwise.

\subsection{Finite-depth two-layer interface}
\label{sec:setup_interface}

The computational domain is
\[
0 \le x \le L_x,\qquad 0 \le y \le L_y,\qquad 0 \le z \le L_z,
\]
with
\[
L_x=164.3674,\qquad L_y=82.1837,\qquad L_z=164.3674 .
\]
The horizontal directions are periodic. The vertical direction is bounded, and the top and bottom portions of the domain contain symmetric far-field relaxation zones. The relaxation width is \(24.6551\), or \(0.15L_z\), so the central interface is separated from each relaxation zone by a substantial interior region during the clean-growth window.

The relaxation zones use a smooth cosine mask that is nonzero only within the top and bottom layers of width \(24.6551\). Within those zones, velocity is nudged toward zero and the scalars are nudged toward the corresponding far-field two-layer target profiles. The relaxation time scale is \(5.0\). This treatment is not an open-boundary model; it is a finite-depth model with remote layers above and below the interface. For that reason, the analysis uses an interior-comparison window before the first standard contact with the relaxation zones and treats later penetration as boundary approach.

The unperturbed interface is centered at \(z_0=L_z/2\). The initial scalar profiles are smooth two-layer jumps displaced by a prescribed horizontal interface height \(\eta(x,y)\):
\begin{align}
  T(x,y,z,0)
    &= \frac{\Delta T}{2}
       \tanh\!\left(\frac{z-z_0-\eta(x,y)}{\delta}\right)
       + T_{\mathrm{seed}}, \\
  S(x,y,z,0)
    &= \frac{\Delta S}{2}
       \tanh\!\left(\frac{z-z_0-\eta(x,y)}{\delta}\right)
       + S_{\mathrm{seed}},
\end{align}
with interface thickness \(\delta=3\). The displacement amplitude is \(A=0.70\), and the scalar/velocity seeds are localized around the interface with nondimensional noise amplitude \(0.04\). The seed terms use the same spectral family as the interface displacement, so the initial roughness and near-interface three-dimensional perturbation are spectrally consistent. The temperature seed amplitude is smaller than the salinity seed amplitude, so the initial perturbation preserves the sharper salinity structure expected in a salt-fingering regime.

\subsection{Controlled spectra and run matrix}
\label{sec:setup_spectra}

The roughness spectra are built from finite sums of oblique horizontal modes with decorrelated phases. Let \(\mathcal{L}(x,y)\) denote the low-mode family, whose selected mode components span \(5 \le k_x \le 16\) and \(2 \le |k_y| \le 8\). Let \(\mathcal{H}(x,y)\) denote the high-annulus family, whose selected mode components span \(7 \le k_x \le 24\) and \(3 \le |k_y| \le 12\). The three primary simulations use
\[
  \eta(x,y) = A\,\mathcal{L}(x,y), \qquad
  \eta(x,y) = A\,\mathcal{H}(x,y), \qquad
  \eta(x,y) = A\,[0.2\mathcal{L}(x,y)+0.8\mathcal{H}(x,y)] .
\]
The fourth simulation repeats the mixed-spectrum forcing with an independent phase and noise realization. \Cref{tab:setup_run_matrix} summarizes the simulation family using descriptive labels.

\begin{table}[tbp]
\centering
\caption{Matched simulation family. All cases use the same physical
parameters, domain, grid, vertical relaxation treatment, time step, and
analysis definitions. The controlled changes are the imposed interfacial
spectrum and, for the replicate, the phase and noise realization.}
\label{tab:setup_run_matrix}
\begin{tabularx}{\linewidth}{>{\raggedright\arraybackslash}p{0.24\linewidth}>{\raggedright\arraybackslash}p{0.23\linewidth}>{\centering\arraybackslash}p{0.11\linewidth}>{\raggedright\arraybackslash}X}
\toprule
Case & Interface spectrum & Amplitude & Purpose \\
\midrule
Mixed spectrum & \(0.2\mathcal{L}+0.8\mathcal{H}\) & 0.70
  & Baseline route-selection case \\
High-annulus spectrum & \(\mathcal{H}\) & 0.70
  & Short-scale branch-locking endpoint \\
Low-mode spectrum & \(\mathcal{L}\) & 0.70
  & Broad-branch transport endpoint \\
Mixed-spectrum replicate & \(0.2\mathcal{L}+0.8\mathcal{H}\) & 0.70
  & Realization robustness check \\
\bottomrule
\end{tabularx}
\end{table}

\subsection{Numerical discretization and saved fields}
\label{sec:setup_numerics}

The simulations are performed with Oceananigans, a finite-volume framework for high-resolution geophysical fluid dynamics
\citep{ramadhan2020oceananigans,wagner2025oceananigans}. The grid
contains \(384 \times 192 \times 960\) cells. With the domain above, this gives horizontal spacings \(\Delta x=\Delta y \approx 0.428\) and vertical spacing \(\Delta z \approx 0.171\). The time step is \(\Delta t=0.0015\), and each calculation is integrated to \(t=60\).

The production comparisons use the nonhydrostatic finite-volume model with centered second-order advection, explicit scalar diffusivities \(\kappa_T=1\) and \(\kappa_S=\tau\), and viscosity \(\nu=\mathrm{Pr}\). Spatial derivatives are evaluated on the staggered rectilinear finite-volume grid, and a nonhydrostatic pressure projection enforces \(\nabla\cdot\boldsymbol{u}=0\) at each time step. No subgrid closure is used beyond the specified molecular viscosity and scalar diffusivities. The evolved fields \((u,v,w,T,S)\) are saved as full three-dimensional outputs and as time-resolved mid-plane slices at nondimensional interval \(0.25\). Fixed point probes are sampled every two time steps at interface, near-plume, and far-field locations. The analysis uses full-volume measures, mid-plane fields, and interior vertically integrated measures that exclude the vertical relaxation zones, so that route selection is not inferred from a single view. The mixed-spectrum replicate is analyzed with the same workflow as the baseline mixed simulation.

The vertical relaxation zones are part of the finite-depth problem rather than open boundaries. The results therefore distinguish a common interior-comparison time, \(t=45\), from later boundary approach. The low-mode case first reaches the standard contact threshold at \(t=50.5\) for vertical velocity and \(t=51.5\) for salinity; later values are retained for finite-depth analysis but not used for the primary three-spectrum interior comparison.

\subsection{Analysis measures and resolution checks}
\label{sec:setup_diagnostics}

The primary measures are chosen to separate morphology, route selection, transport, vertical reach, and scalar-interface geometry. Unless otherwise specified, \(\langle \cdot\rangle_{xy}\) denotes a horizontal average, \(\langle \cdot\rangle_i\) denotes a volume average over the interior region between the vertical relaxation zones, and primes denote departures from the corresponding horizontal mean.

For a mid-plane field \(q(x,y,t)\), the horizontal Fourier coefficient is \(\widehat{q}_{mn}(t)\). The modal power is
\begin{equation}
  P_{mn}^{(q)}(t)=|\widehat{q}_{mn}(t)|^2 ,
\end{equation}
with the zero mode removed. Radial spectra are formed by summing this power over annuli in horizontal wavenumber. The normalized shell fraction is
\begin{equation}
  E_j^{(q)}(t)
  =
  \frac{\sum_{\boldsymbol{k}_{mn}\in \mathcal{A}_j}
  P_{mn}^{(q)}(t)}
  {\sum_{\boldsymbol{k}_{mn}}P_{mn}^{(q)}(t)} ,
\end{equation}
where \(\mathcal{A}_j\) is the \(j\)-th radial annulus. The dominant wavelength is obtained from the radial bin or discrete mode carrying the largest power. The effective modal population is a participation ratio,
\begin{equation}
  N_{\mathrm{eff}}^{(q)}
  =
  \frac{\left(\sum_{\boldsymbol{k}}P_{\boldsymbol{k}}^{(q)}\right)^2}
       {\sum_{\boldsymbol{k}}\left(P_{\boldsymbol{k}}^{(q)}\right)^2}.
  \label{eq:effective_modes}
\end{equation}
This definition equals one for a single active mode and increases as power is distributed over more planform modes.

Route handoff is measured by branch-power ratios. If \(B\) is the tracked broad branch and \(I\) is the initially imposed branch, then
\begin{equation}
  \mathcal{R}^{(q)}(t)
  =
  \frac{\sum_{\boldsymbol{k}\in B}P_{\boldsymbol{k}}^{(q)}(t)}
       {\sum_{\boldsymbol{k}\in I}P_{\boldsymbol{k}}^{(q)}(t)} .
  \label{eq:branch_ratio}
\end{equation}
The reported handoff time is the first persistent crossing of \(\mathcal{R}^{(q)}=1\). We evaluate this in the mid-plane and in an interior vertically integrated measure so that branch selection is not inferred from a single slice.

Angular organization is measured from the nematic moment of the Fourier power distribution,
\begin{equation}
  \mathcal{A}^{(q)}
  =
  \left|
  \frac{\sum_{\boldsymbol{k}} P_{\boldsymbol{k}}^{(q)}
  \exp(2i\theta_{\boldsymbol{k}})}
       {\sum_{\boldsymbol{k}}P_{\boldsymbol{k}}^{(q)}}
  \right|,
  \qquad
  \theta_p^{(q)}
  =
  \frac{1}{2}\arg\!
  \left[
  \frac{\sum_{\boldsymbol{k}} P_{\boldsymbol{k}}^{(q)}
  \exp(2i\theta_{\boldsymbol{k}})}
       {\sum_{\boldsymbol{k}}P_{\boldsymbol{k}}^{(q)}}
  \right],
  \label{eq:angular_anisotropy}
\end{equation}
where \(\theta_{\boldsymbol{k}}\) is the horizontal wavevector angle. The factor of two folds opposite directions onto the same axis. Signed-branch content distinguishes the two oblique orientations of a branch. For a tracked mode \((m,n)\), the same-sign and opposite-sign powers are
\begin{align}
  P_{\mathrm{same}}^{(q)}
    &= P_{m,n}^{(q)} + P_{-m,-n}^{(q)}, \\
  P_{\mathrm{opp}}^{(q)}
    &= P_{m,-n}^{(q)} + P_{-m,n}^{(q)} ,
\end{align}
and the signed orientation bias is
\begin{equation}
  \mathcal{B}^{(q)}
  =
  \frac{P_{\mathrm{same}}^{(q)}-P_{\mathrm{opp}}^{(q)}}
       {P_{\mathrm{same}}^{(q)}+P_{\mathrm{opp}}^{(q)}} .
  \label{eq:signed_branch_bias}
\end{equation}

The primary transport metric is the interior down-gradient salinity flux,
\begin{equation}
  F_S(t)=-\langle w S'\rangle_i .
  \label{eq:salt_flux}
\end{equation}
Supporting transport and activity measures include
\begin{equation}
  K=\frac{1}{2}\langle u^2+v^2+w^2\rangle_i,\qquad
  w_{\mathrm{rms}}=\langle w^2\rangle_i^{1/2},\qquad
  \sigma_S^2=\langle (S')^2\rangle_i .
  \label{eq:transport_support}
\end{equation}
For local plume-passage and upper/lower balance, asymmetry is reported as
\begin{equation}
  \mathcal{Q}_{\mathrm{asym}}
  =
  \frac{Q_{\mathrm{upper}}-Q_{\mathrm{lower}}}
       {Q_{\mathrm{upper}}+Q_{\mathrm{lower}}},
  \label{eq:upper_lower_asymmetry}
\end{equation}
so negative values denote lower-biased activity.

Vertical reach is measured with activity envelopes. For \(w\), \(a_w(z,t)=\langle w^2\rangle_{xy}^{1/2}\); for salinity, \(a_S(z,t)\) is the horizontal standard deviation of \(S\). Given a threshold \(\alpha\), the active set is
\begin{equation}
  \mathcal{Z}_{q,\alpha}(t)
  =
  \left\{z:\; a_q(z,t)\ge \alpha\max_z a_q(z,t)\right\}.
  \label{eq:activity_set}
\end{equation}
The active width is
\begin{equation}
  W_{q,\alpha}(t)=
  \max \mathcal{Z}_{q,\alpha}(t)-\min \mathcal{Z}_{q,\alpha}(t),
  \label{eq:activity_width}
\end{equation}
and the minimum distance from the active set to the inner edges of the relaxation zones is used to define the first boundary-region arrival time. The standard threshold used for the reported contact times is \(\alpha=0.3\), with threshold sensitivity reported separately.

Interface geometry is measured in two ways. The temperature-zero surface \(h_T(x,y,t)\) is the zero crossing of \(T\) closest to the original interface height \(z_0\). Its RMS displacement is
\begin{equation}
  h_{T,\mathrm{rms}}(t)
  =
  \left\langle \left[h_T(x,y,t)-z_0\right]^2
  \right\rangle_{xy}^{1/2}.
  \label{eq:temperature_zero_height}
\end{equation}
For the active salinity interface, the vertical gradient weight is \(G=|\partial_z S|^2\) within a window surrounding \(z_0\). The gradient-centroid height and local gradient thickness are
\begin{align}
  h_{\nabla S}(x,y,t)
    &=
    \frac{\int z\,G(x,y,z,t)\,dz}
         {\int G(x,y,z,t)\,dz}, \\
  d_{\nabla S}(x,y,t)
    &=
    \left[
    \frac{\int (z-h_{\nabla S})^2 G(x,y,z,t)\,dz}
         {\int G(x,y,z,t)\,dz}
    \right]^{1/2}.
  \label{eq:salinity_gradient_interface}
\end{align}
The reported salinity-gradient effective mode count is computed from \(h_{\nabla S}\) using \cref{eq:effective_modes}.

The simulations resolve the analysis scales used in the comparison. Energetic spectral content is separated from the grid cutoff, the time-step measure remains small, and the dominant planform wavelengths span many grid cells. These checks support the modal route selection, finite-depth reach, and interface geometry measured over the simulated interval. The study does not attempt to infer a universal grid-converged flux law.

The literature comparison used here is qualitative rather than a one-to-one benchmark against a homogeneous fingering state. The parameters place the calculations inside the classical salt-fingering regime
\citep{stern1960salt,radko2013double}. The density-ratio language and
oceanographic use follow standard double-diffusive practice
\citep{ruddick1983practical,schmitt1994double}. The observed progression from
organized fingers toward three-dimensional, secondary structure is consistent with finite-length and secondary-instability theory
\citep{holyer1984stability,kunze1987limits}. It is also consistent with
canonical nonlinear simulations in which salt fingers progress toward chaotic convection \citep{simeonov2009dns}. The transport measures follow the convention that the down-gradient salinity flux is the primary scalar-exchange measure in fingering convection \citep{traxler2011dynamics}. The interface and vertical-envelope measures distinguish plume-forest reorganization from thermohaline staircase formation, whose mean-field and observational settings are related but not identical to the present finite-depth rough-interface problem
\citep{merryfield2000origin,radko2003mechanism,stellmach2011dynamics}.

\section{Results}
\label{sec:results}

The results are organized around two intervals in the life of a finite-depth thermohaline interface. The first is an interior-growth window, during which the spectra can be compared before the most rapidly penetrating plume forest reaches the relaxation zones. The second is a finite-depth boundary-approach window, during which the selected routes separate according to how efficiently they connect the original interface to remote layers. We use \(t=45\) as the strict common interior-comparison time, because the low-mode spectrum does not reach the standard relaxation-zone contact threshold until \(t=50.5\) in the vertical-velocity envelope and \(t=51.5\) in the salinity envelope. Values at later times describe finite-depth reach and boundary approach; they are not treated as equally interior comparisons for all spectra. All quantities reported in this section are nondimensional.

The central result is that the imposed roughness spectrum controls the route to salt-finger mixing beyond the amount of disturbance energy injected into the initial condition. The high-annulus spectrum remains compact and locked to its imposed short-scale branch. The low-mode spectrum immediately selects a broad branch, producing the strongest interior-comparison transport and the earliest finite-depth reach. The mixed spectrum follows a third route, in which vertical velocity organizes onto the broad branch before salinity, while the salinity field develops the richest scalar spectral population.

Here, route refers to the coupled evolution of organizing wavelength, branch topology, scalar spectral population, transport, active-interface geometry, and vertical reach. This definition combines multiple measures because a spectrum can create a visually recognizable finger forest without maximizing transport, and a strongly transporting route can have a relatively narrow scalar spectral population. The Results follow the physics sequence. We first show how the spectrum selects a route, then how that route changes transport and scalar spectral richness, how it reshapes the active scalar interface, how far the plume forest reaches over the finite depth, and whether the mixed route survives a second realization.

\begin{table}[tbp]
\centering
\caption{Summary of the physical claims and supporting measures.}
\label{tab:main_claim_evidence_map}
\begin{tabularx}{\linewidth}{>{\raggedright\arraybackslash}p{0.24\linewidth}>{\raggedright\arraybackslash}X>{\raggedright\arraybackslash}p{0.24\linewidth}}
\toprule
Claim & Supporting measures & Physical meaning \\
\midrule
Spectrum selects route & Matched morphology, route-crossing table, mechanism metrics, and finite-depth reach. & The spectrum changes both disturbance amplitude and the nonlinear pathway. \\
High-annulus forcing stays compact & No broad-branch crossing by \(t=60\), \(F_S=0.05642\) at \(t=45\), \(S\) effective modes \(=3.26\). & Short-scale roughness can remain branch-locked and weakly transporting. \\
Low-mode forcing transports strongest & \(F_S=0.20068\), \(w_{\rm rms}=0.86263\), and \(w\)-activity width \(80.13\) at \(t=45\). & Broad initial roughness gives the strongest interior exchange and earliest finite-depth reach. \\
Mixed forcing is route-distinct & Mixed \(w\) crosses before salinity; \(S\) effective modes \(=86.66\) at \(t=45\) and \(159.88\) at \(t=57.75\). & The mixed response is a distinct velocity-led, scalar-rich route. \\
Seed replicate preserves route & Independent mixed realization retains positive \(S-w\) lag, broad \(S\) wavelength, high \(S\) effective modes, and same-order transport. & The mixed route is robust in ordering and measured response family, while exact timing is realization-dependent. \\
Resolution checks support the route comparison & Small CFL, weak high-wavenumber tails, and dominant wavelengths separated from the grid cutoff. & The tracked route-selection scales are resolved over the analyzed window. \\
\bottomrule
\end{tabularx}
\end{table}

\Cref{tab:main_claim_evidence_map} maps the main physics claims to their
supporting measures. These measures separate three routes rather than ranking a single ``best spectrum'': compact high-annulus branch locking, immediate low-mode penetration, and delayed mixed velocity-led reorganization with a strongly populated scalar spectrum. The seed replicate tests whether the mixed route survives a second phase/noise realization. This organization is useful because the most vigorous transport route is not the route with the richest scalar spectrum, and the visually most complex plume forest is not the first one to reach the finite-depth boundary region.

\subsection{Route selection by the imposed interfacial spectrum}
\label{sec:results_route_morphology}

\begin{figure}[tbp]
  \centering
  \includegraphics[width=\linewidth]{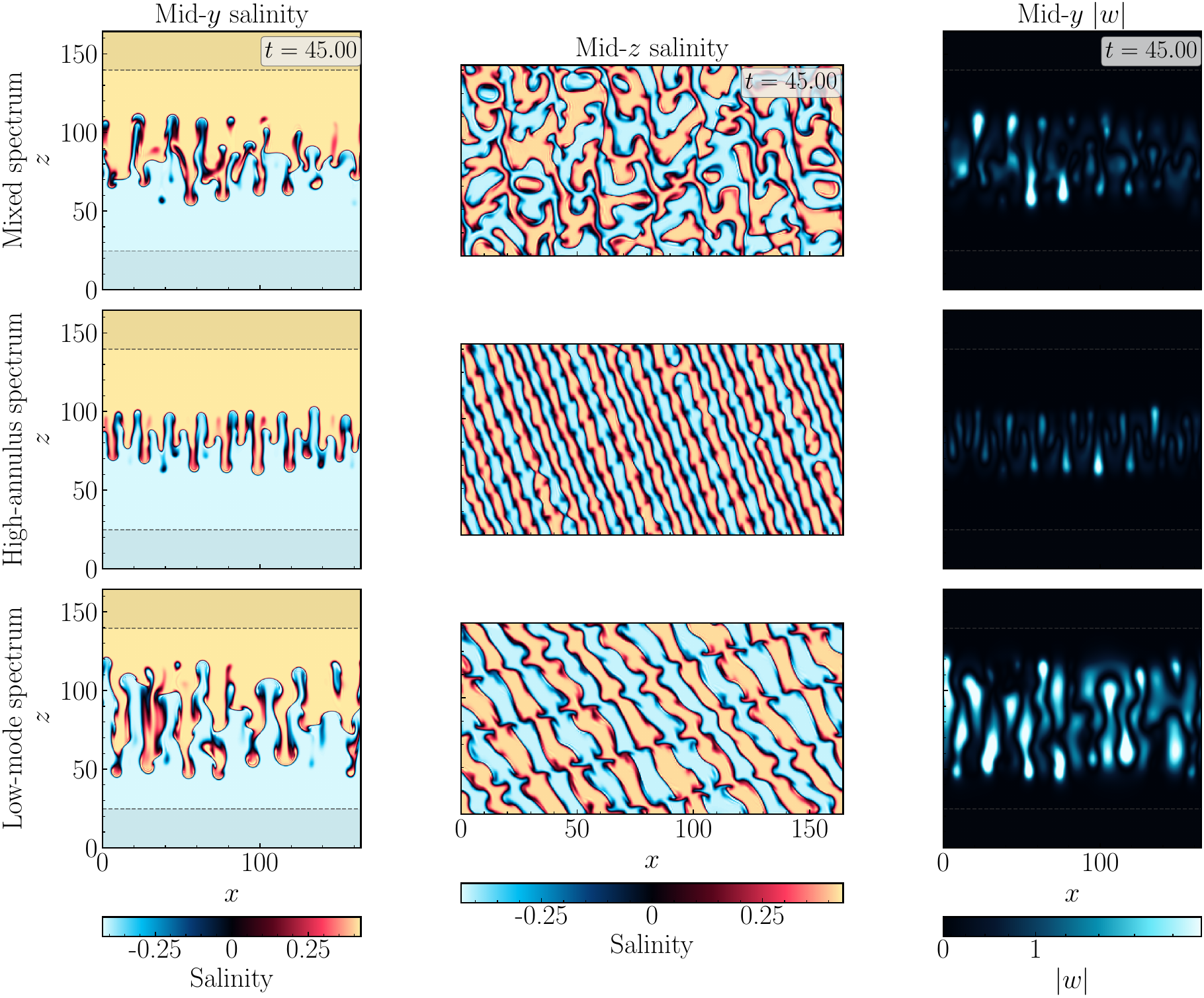}
  \caption{Interior-comparison plume-forest morphology at \(t=45\) for the three
  imposed spectra. All cases use matched physical parameters, domain, grid,
  relaxation-zone treatment, time step, and output cadence. Only the
  imposed interfacial spectrum differs. Color limits are matched across cases
  for each displayed field and plane. The comparison time precedes the first
  standard low-mode relaxation-zone contacts, which occur at \(t=50.5\) for
  vertical velocity and \(t=51.5\) for salinity.}
  \label{fig:three_case_morphology_clean_t45}
\end{figure}

\Cref{fig:three_case_morphology_clean_t45} compares the three imposed
roughness spectra at \(t=45\). At this common interior time, the plume forests are already visibly distinct. The high-annulus case remains compact and localized near the original interface. The low-mode case forms broad vertically penetrating structures that occupy a much larger fraction of the interior. The mixed case develops a richer scalar texture and broader organization than the high-annulus endpoint, without reducing to an average of the high-annulus and low-mode responses.

The morphology comparison establishes that the controlled spectral perturbation has selected different nonlinear states before finite-depth relaxation-zone interaction becomes important. Transport ranking comes from the route, transport, spectral-population, and penetration measures that follow. Because the physical parameters, domain, grid, relaxation-zone treatment, and comparison times are matched, the visual differences at \(t=45\) reflect route differences rather than plotting or sampling differences.

The high-annulus morphology is the first warning that finer imposed roughness does not necessarily produce a finer, more active salt-finger forest. The short-scale branch remains visible, but the active region stays comparatively confined. The low-mode morphology shows broad structures that develop early and extend far from the initial interface. The mixed case is visually important because it contains broad organization and scalar texture at the same time. This combination identifies mixed roughness as a distinct route, not a midpoint between the two single-band spectra. In physical terms, the roughness spectrum does more than set the initial finger width. It steers how the unstable interface organizes the exchange pathway that develops from that roughness.

Thus the morphology already identifies locked compact growth, immediate broad penetration, and delayed mixed-route reorganization before the modal, transport, and interface metrics are introduced. The high-annulus case falsifies the simple expectation that smaller imposed scales necessarily create a finer and more effective plume forest. It forms fingers, but the active region remains closer to the initial interface and the planform remains tied to the imposed short branch. The low-mode case is broad from the beginning, while the mixed case contains both broad organization and fine scalar texture.

\subsection{Modal handoff, branch topology, and plume-passage structure}
\label{sec:results_route_selection}

The route-selection measures make the morphology quantitative. We track the relative strength of the initially imposed branch and the broad branch in both a mid-plane measure and an interior vertically integrated measure. The mid-plane measure is sensitive to the view used in the morphology figures. The interior integrated measure excludes the vertical relaxation zones and checks that the same transition is present in the interior volume instead of appearing only in one slice.

\Cref{tab:route_selection} shows that the high-annulus case remains
locked to the imposed \((16,3)\) branch through \(t=60\), with no detected broad-branch crossing in either vertical velocity or salinity. This branch locking shows that the high-wavenumber perturbation does not quickly coarsen into the tracked broad branch under the present finite-depth conditions. The low-mode case behaves as the opposite endpoint. The broad branch is present essentially from the start, with interior-integrated salinity already broad at \(t=0\), interior-integrated vertical velocity crossing at \(t=1.0\), and mid-plane vertical velocity crossing at \(t=2.0\). It begins on the broad route rather than undergoing a delayed handoff.

\begin{table}[tbp]
\centering
\caption{Route-selection measures for the three imposed spectra. Crossing times are nondimensional. The mixed case shows delayed velocity-led handoff, high-annulus forcing has no detected broad-branch crossing by \(t=60\), and low-mode forcing is broad almost immediately. Interior crossings exclude the outer relaxation layers; outer-zone arrival gives first threshold contact with those layers. ``n/a'' indicates no detected crossing or arrival by \(t=60\); in the low-mode salinity handoff column, the interior measure is already broad at the earliest sampled time, as shown by the interior \(S\) value of \(0\).}
\label{tab:route_selection}
\begin{tabular}{lcccc}
\toprule
Case & \(w\) handoff & \(S\) handoff & Interior \(w/S\) & Outer-zone \(w/S\) \\
\midrule
Mixed spectrum & 36.5 & 44.75 & 39.5/47 & 57.75/59.5 \\
High-annulus spectrum & n/a & n/a & n/a/n/a & n/a/n/a \\
Low-mode spectrum & 2 & n/a & 1/0 & 50.5/51.5 \\
\bottomrule
\end{tabular}
\end{table}

\begin{figure}[tbp]
  \centering
  \includegraphics[width=\linewidth]{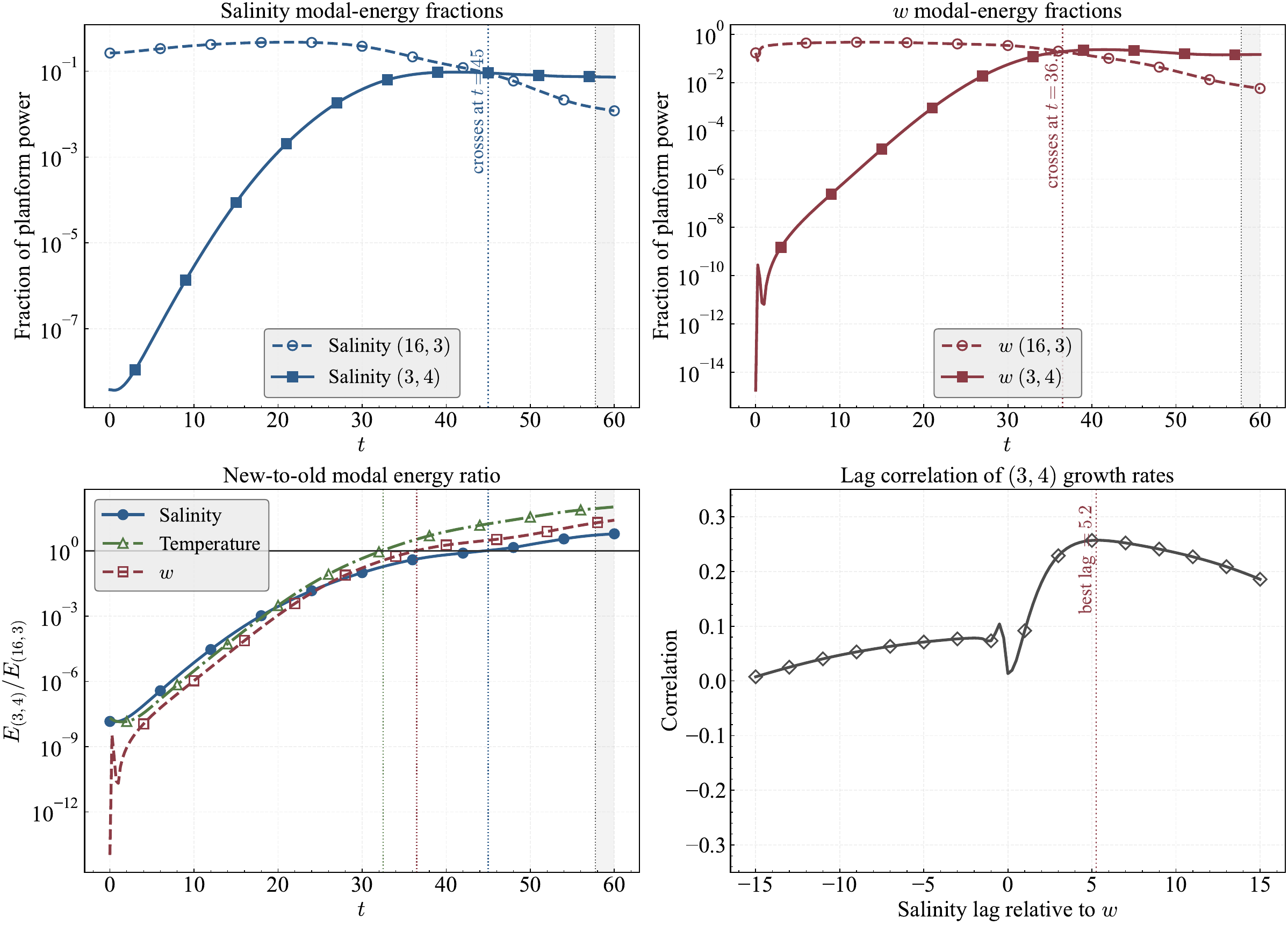}
  \caption{Mixed-spectrum modal-lag measure. The tracked broad-branch
  ratio crosses first in the velocity field and later in salinity, so the
  mixed route is velocity-led in timing. The same ordering is checked in the
  interior vertically integrated measure reported in
  \cref{tab:route_selection}.}
  \label{fig:modal_lag_main}
\end{figure}

\begin{figure}[tbp]
  \centering
  \includegraphics[width=\linewidth]{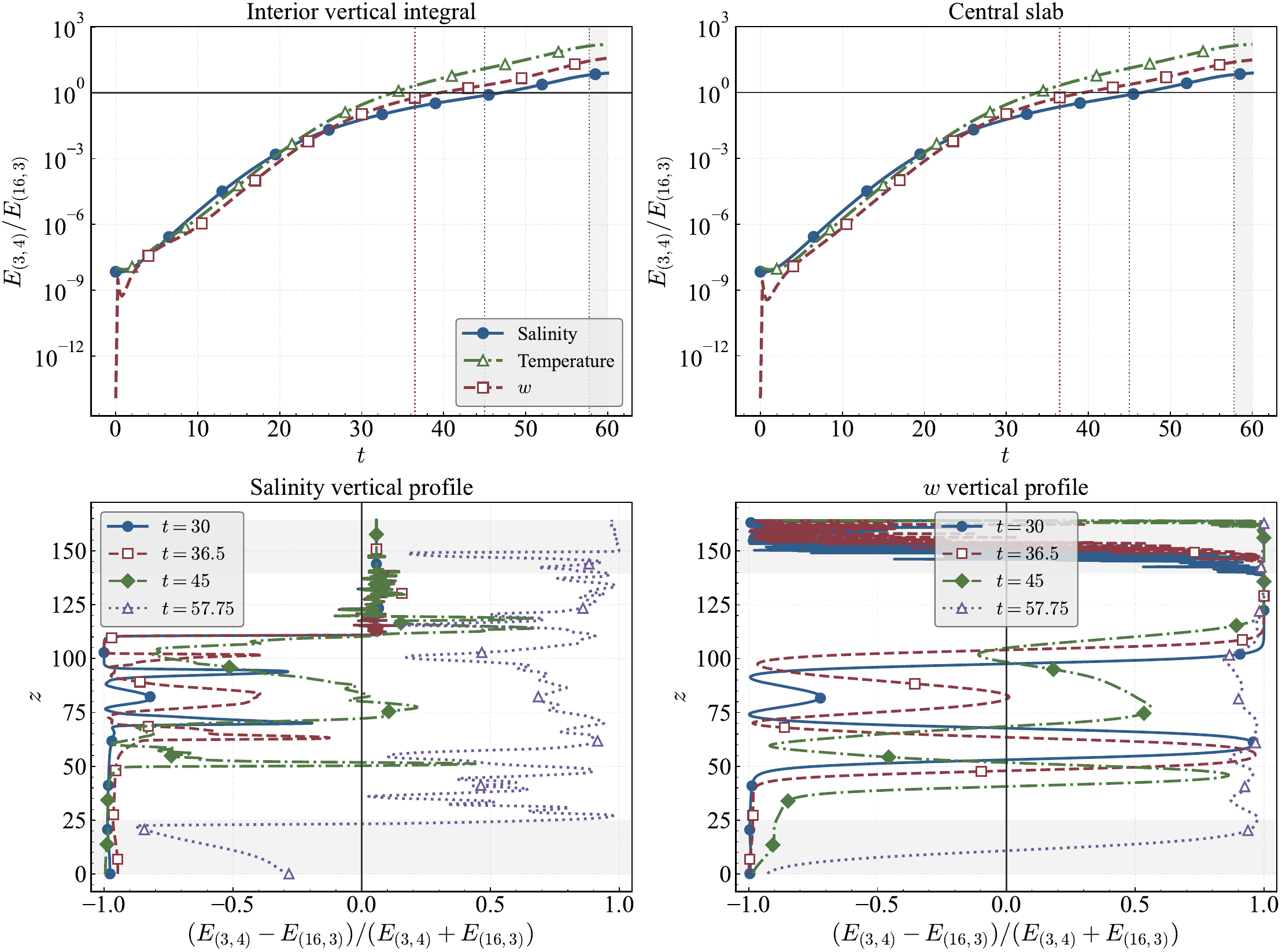}
  \caption{Cross-plane and interior-integrated modal checks for the
  mixed-spectrum route. The velocity-led broad-branch transition is present in
  the interior-volume measure as well as the mid-plane measure, showing
  that the route-selection result is not a single-slice artifact.}
  \label{fig:modal_crossplane_integrated}
\end{figure}

The mixed case exhibits a delayed velocity-led route (\cref{fig:modal_lag_main}). In the mid-plane, the vertical velocity reaches the tracked broad branch at \(t=36.5\), while salinity follows near \(t=44.75\). The same ordering persists in the interior vertically integrated measure, where vertical velocity crosses at \(t=39.5\) and salinity at \(t=47.0\) (\cref{fig:modal_crossplane_integrated}). This ordering is a route-selection measure. The velocity field reorganizes onto the broad branch before the scalar field fully follows. We therefore describe the mixed case as velocity-led in timing and organization, without implying one-way causality from velocity to salinity.

This ordering is the clearest evidence that the mixed case differs from a linear superposition of high-annulus and low-mode behavior. The high-annulus case does not cross, the low-mode case is broad almost immediately, and the mixed case has a delayed velocity-led transition in both the mid-plane and interior vertically integrated measures. The mid-plane measure connects the result to the visible planform morphology, whereas the interior vertically integrated measure protects the claim from being a single-slice artifact. The transition times are not identical in the two measures, but the ordering is consistent: in the mixed case, the flow field reorganizes onto the broad branch before the scalar field fully follows.

\begin{figure}[tbp]
  \centering
  \begin{subfigure}{0.48\linewidth}
    \centering
    \includegraphics[width=\linewidth]{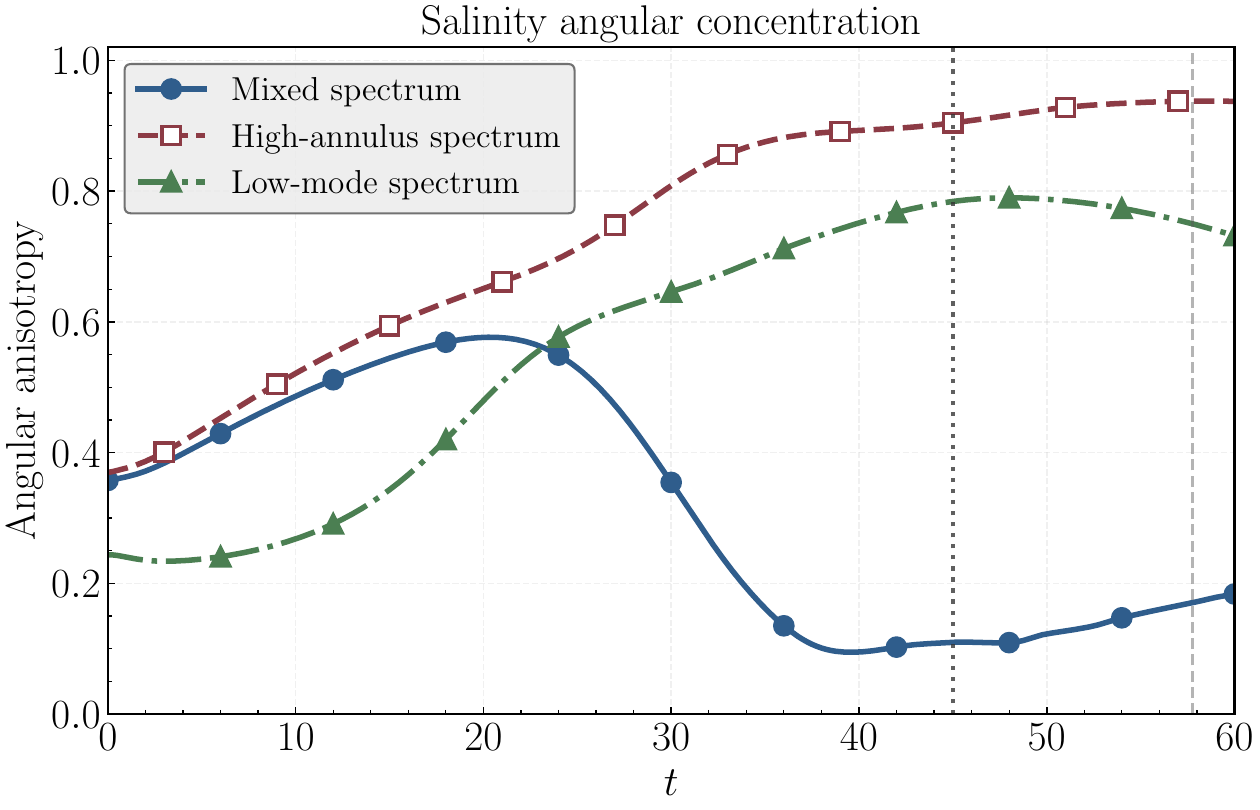}
    \caption{Salinity anisotropy}
    \label{fig:angular_s_anisotropy}
  \end{subfigure}\hfill
  \begin{subfigure}{0.48\linewidth}
    \centering
    \includegraphics[width=\linewidth]{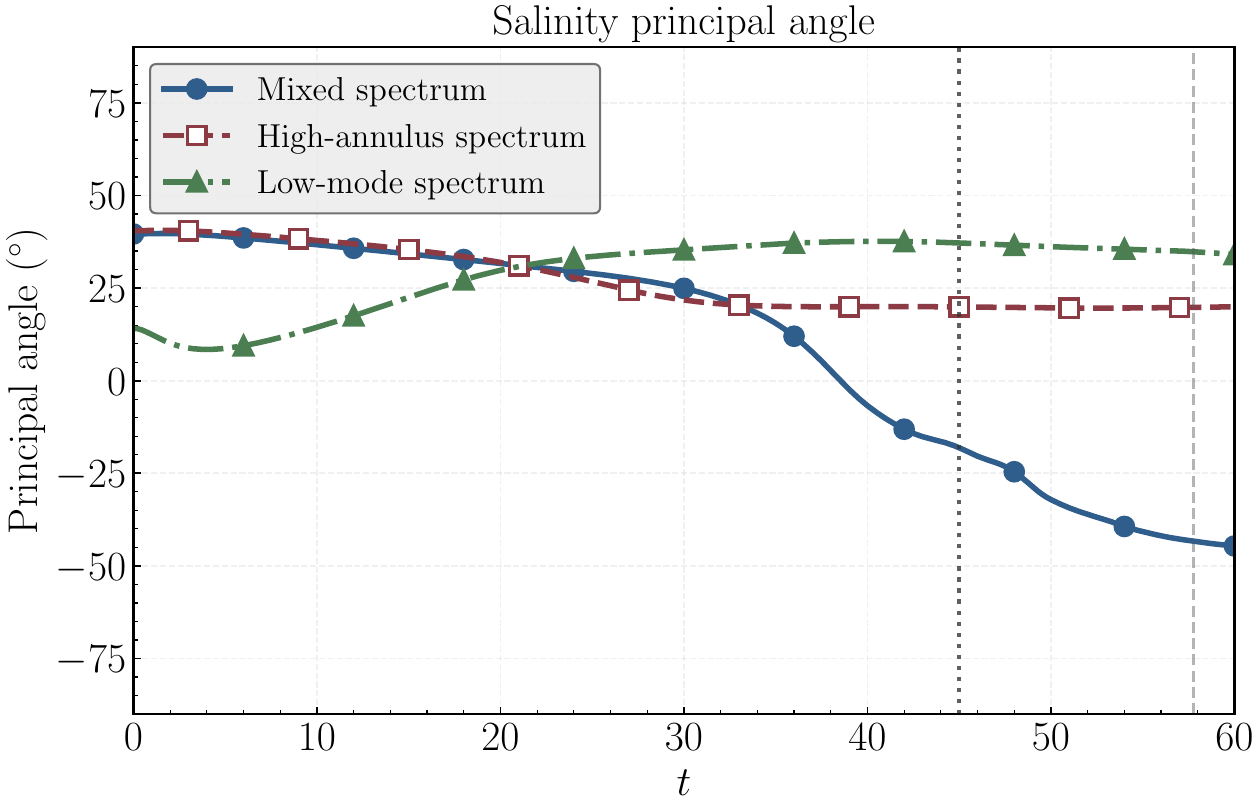}
    \caption{Salinity orientation}
    \label{fig:angular_s_angle}
  \end{subfigure}
  \vspace{0.6em}
  \begin{subfigure}{0.48\linewidth}
    \centering
    \includegraphics[width=\linewidth]{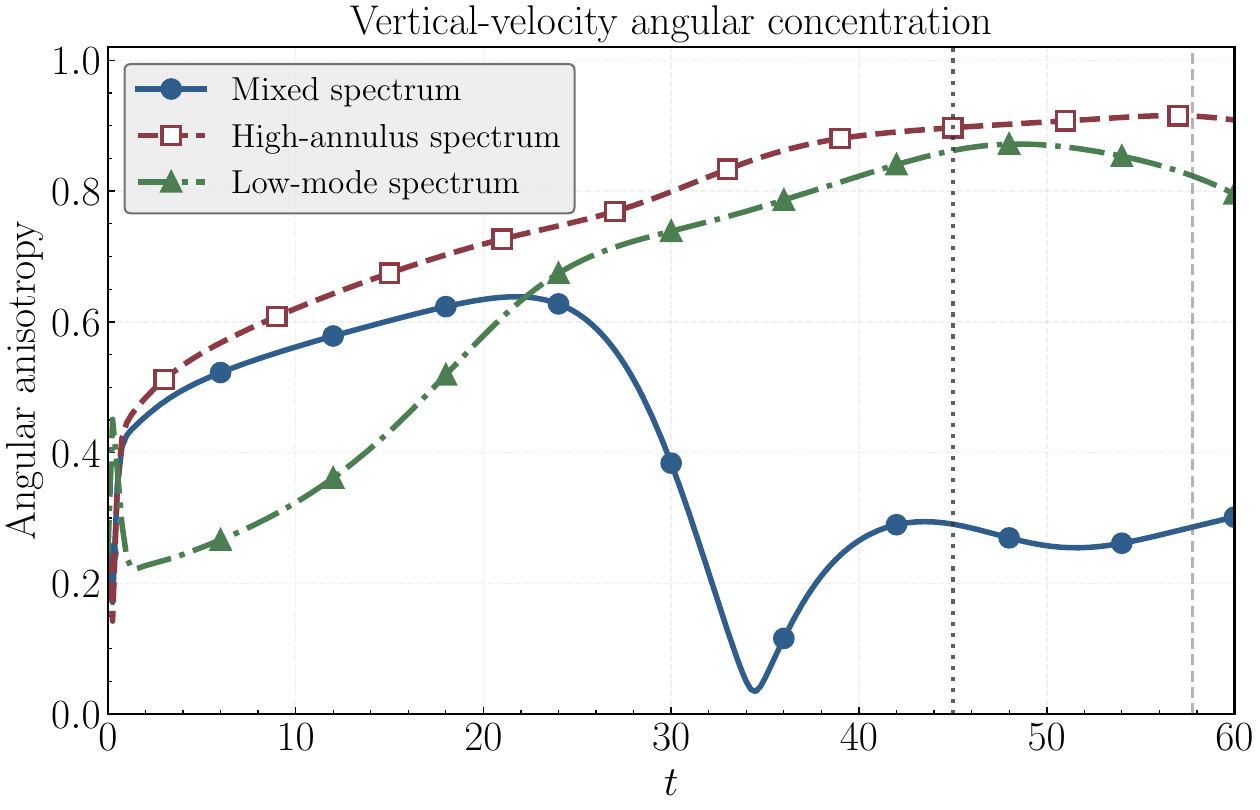}
    \caption{Vertical-velocity anisotropy}
    \label{fig:angular_w_anisotropy}
  \end{subfigure}\hfill
  \begin{subfigure}{0.48\linewidth}
    \centering
    \includegraphics[width=\linewidth]{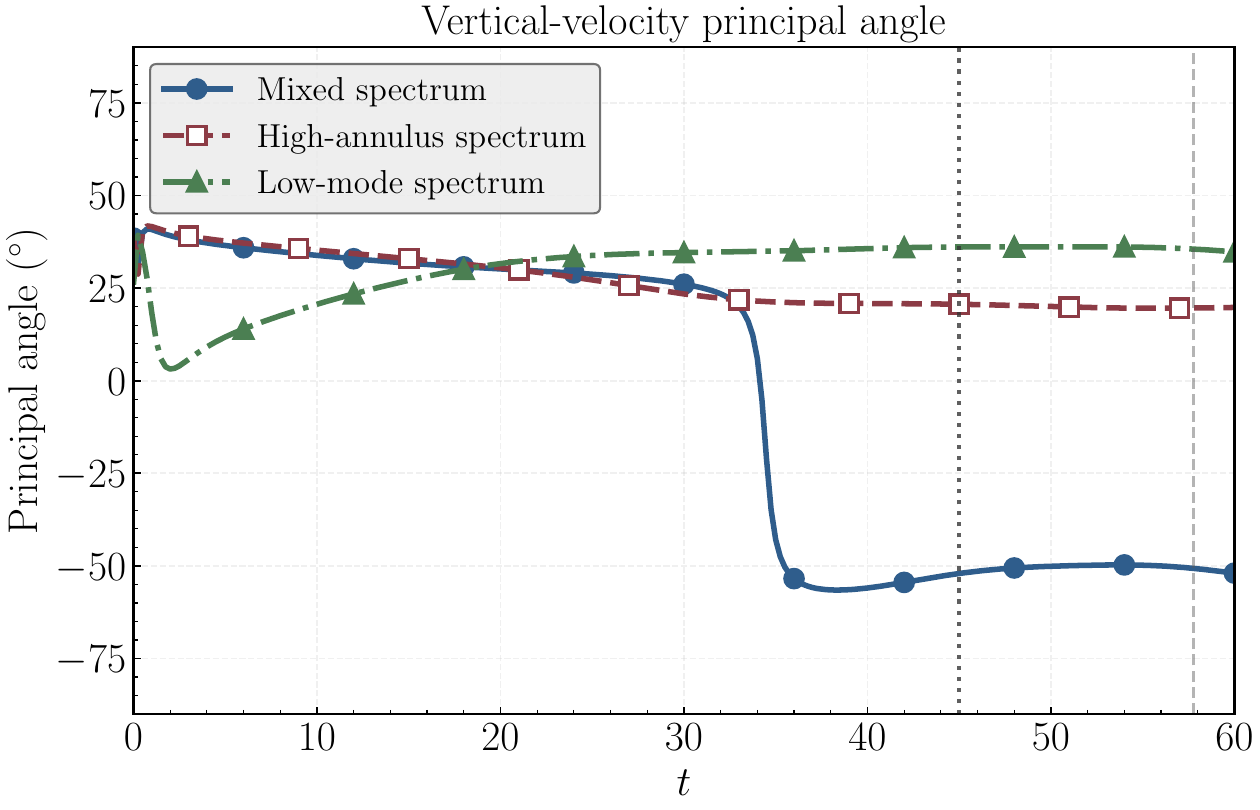}
    \caption{Vertical-velocity orientation}
    \label{fig:angular_w_angle}
  \end{subfigure}
  \caption{Angular organization at the interior-comparison time. The
  high-annulus and low-mode endpoints are strongly oriented on branch-specific
  folded angles, while the mixed case has lower single-angle anisotropy because
  competing signed branches coexist.}
  \label{fig:angular_organization}
\end{figure}

\begin{figure}[tbp]
  \centering
  \begin{subfigure}{0.48\linewidth}
    \centering
    \includegraphics[width=\linewidth]{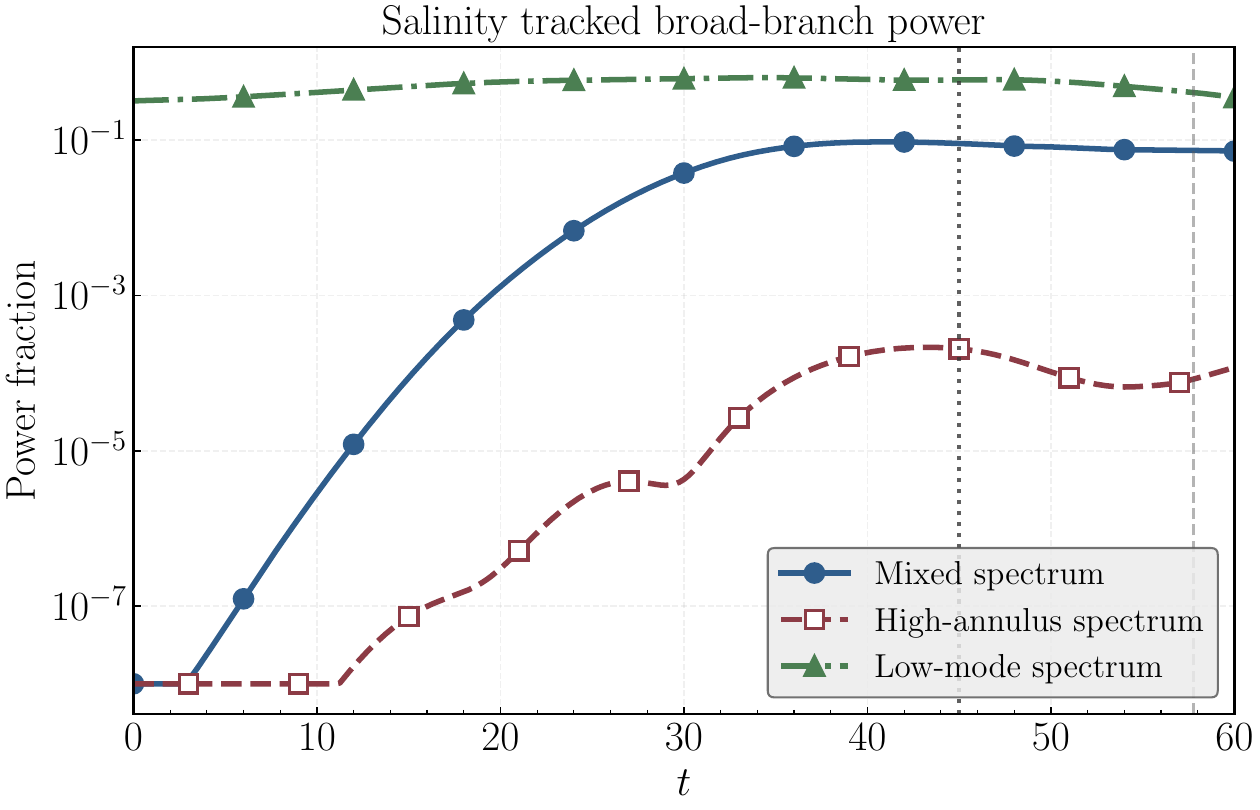}
    \caption{Salinity branch power}
    \label{fig:signed_s_branch_power}
  \end{subfigure}\hfill
  \begin{subfigure}{0.48\linewidth}
    \centering
    \includegraphics[width=\linewidth]{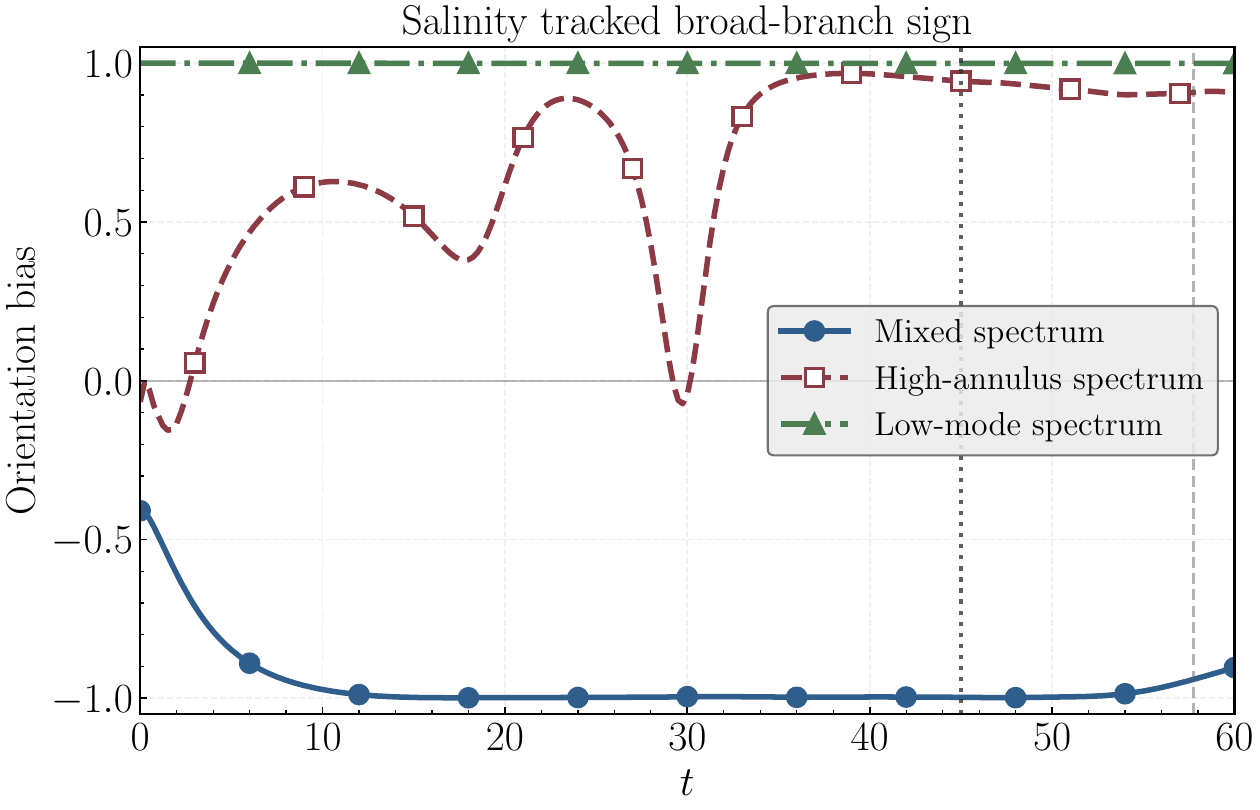}
    \caption{Salinity branch sign}
    \label{fig:signed_s_branch_sign}
  \end{subfigure}
  \vspace{0.6em}
  \begin{subfigure}{0.48\linewidth}
    \centering
    \includegraphics[width=\linewidth]{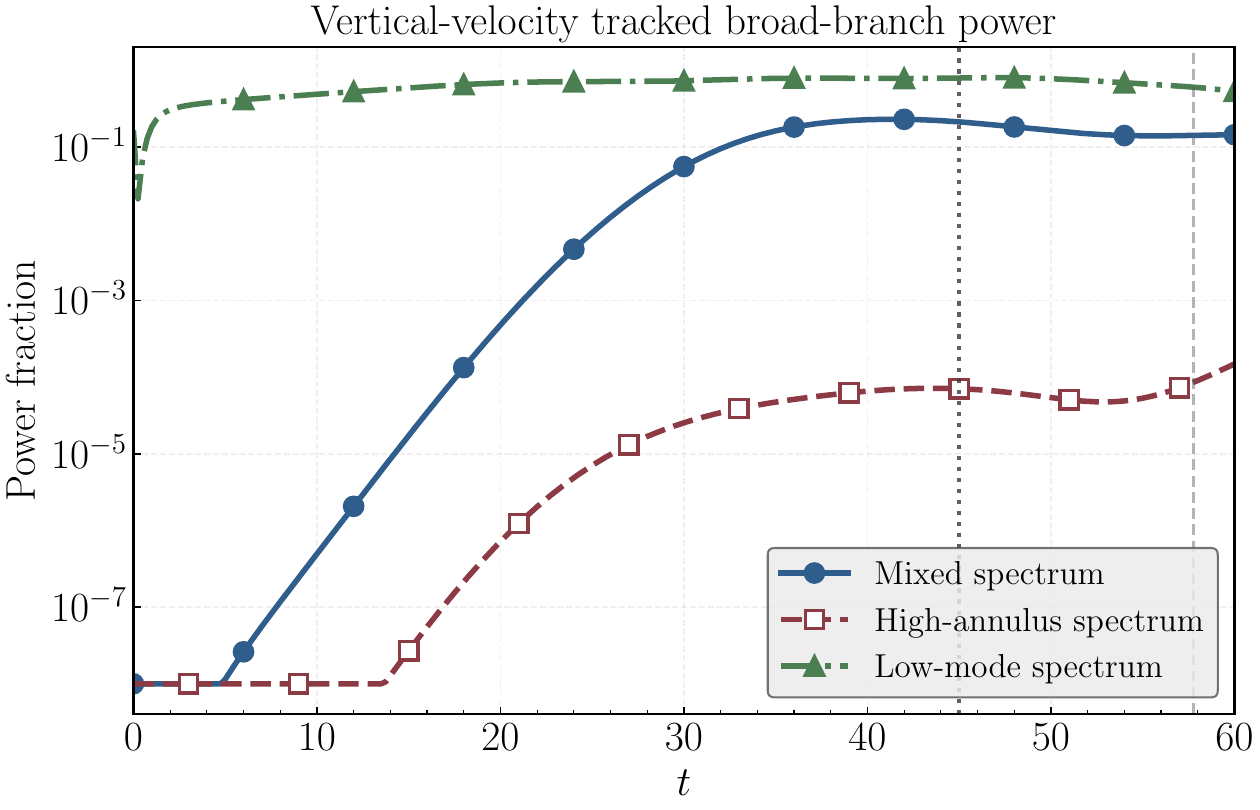}
    \caption{Vertical-velocity branch power}
    \label{fig:signed_w_branch_power}
  \end{subfigure}\hfill
  \begin{subfigure}{0.48\linewidth}
    \centering
    \includegraphics[width=\linewidth]{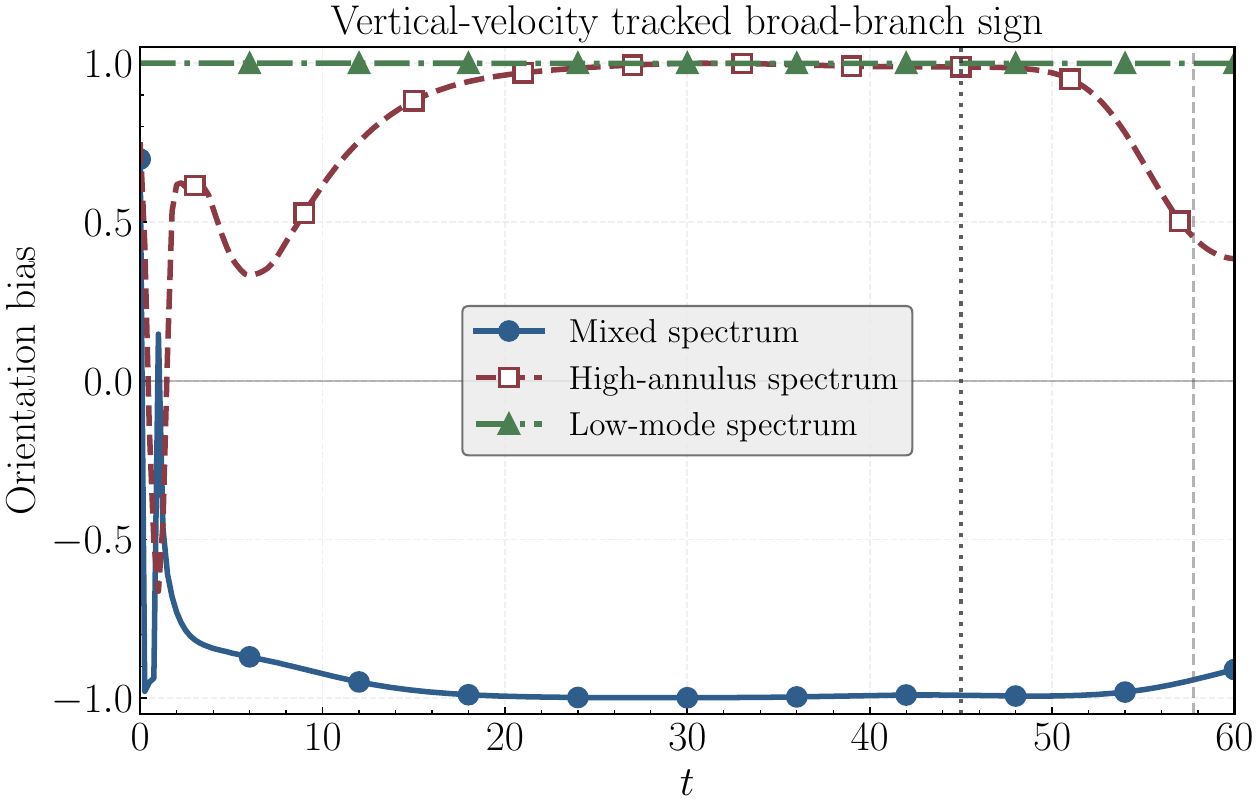}
    \caption{Vertical-velocity branch sign}
    \label{fig:signed_w_branch_sign}
  \end{subfigure}
  \caption{Signed branch measures at the interior-comparison time. The
  endpoint spectra concentrate power on their selected branches, whereas the
  mixed route contains strong opposite-sign broad-branch contributions and a
  broader modal population.}
  \label{fig:signed_branch_diagnostics}
\end{figure}

The radial and branch-crossing measures identify which organizing scale is selected. Angular and signed-branch measures (\cref{fig:angular_organization,fig:signed_branch_diagnostics}) add a complementary part of the same route-selection result by showing how that scale is oriented in the horizontal plane and whether the apparent diagonal or squiggly structures belong to one dominant signed branch or to competing branches. This matters because a low single-angle anisotropy does not necessarily mean a weakly organized scalar field. In the mixed case, it can mean that multiple signed branches coexist while the scalar spectrum is becoming highly populated.

The salinity anisotropy and orientation panels
\cref{fig:angular_s_anisotropy,fig:angular_s_angle}, together with the
corresponding vertical-velocity panels
\cref{fig:angular_w_anisotropy,fig:angular_w_angle}, show that, at \(t=45\),
high-annulus salinity is strongly anisotropic, with anisotropy \(0.904\) at folded angle \(19.9^\circ\), and its vertical velocity is similarly anisotropic, with anisotropy \(0.897\) at \(20.6^\circ\). Low-mode forcing is also strongly oriented on a broader branch. Salinity has anisotropy \(0.784\) at \(37.2^\circ\), while vertical velocity has anisotropy \(0.862\) at \(36.1^\circ\). The mixed case has much smaller single-angle anisotropy, \(0.110\) in salinity and \(0.291\) in vertical velocity, because competing signed branches coexist. The signed-branch panels
\cref{fig:signed_s_branch_power,fig:signed_s_branch_sign,fig:signed_w_branch_power,fig:signed_w_branch_sign}
show that, in the tracked broad branch, the mixed salinity field has branch power fraction \(0.091\) with opposite-sign orientation bias \(-0.998\). Vertical velocity has branch power fraction \(0.214\) with bias \(-0.991\). By contrast, the high-annulus initial branch and low-mode broad branch carry much larger salinity branch power fractions, \(0.781\) and \(0.601\), and vertical-velocity branch power fractions, \(0.849\) and \(0.802\), respectively.

\Cref{fig:angular_organization,fig:signed_branch_diagnostics} show that
diagonal or squiggly organization does not imply the same route in every spectrum. Endpoint fields are anisotropic because one branch dominates; the mixed case has lower folded-angle anisotropy because competing signed branches coexist while the scalar population broadens.

\begin{figure}[tbp]
  \centering
  \begin{subfigure}{0.32\linewidth}
    \centering
    \includegraphics[width=\linewidth]{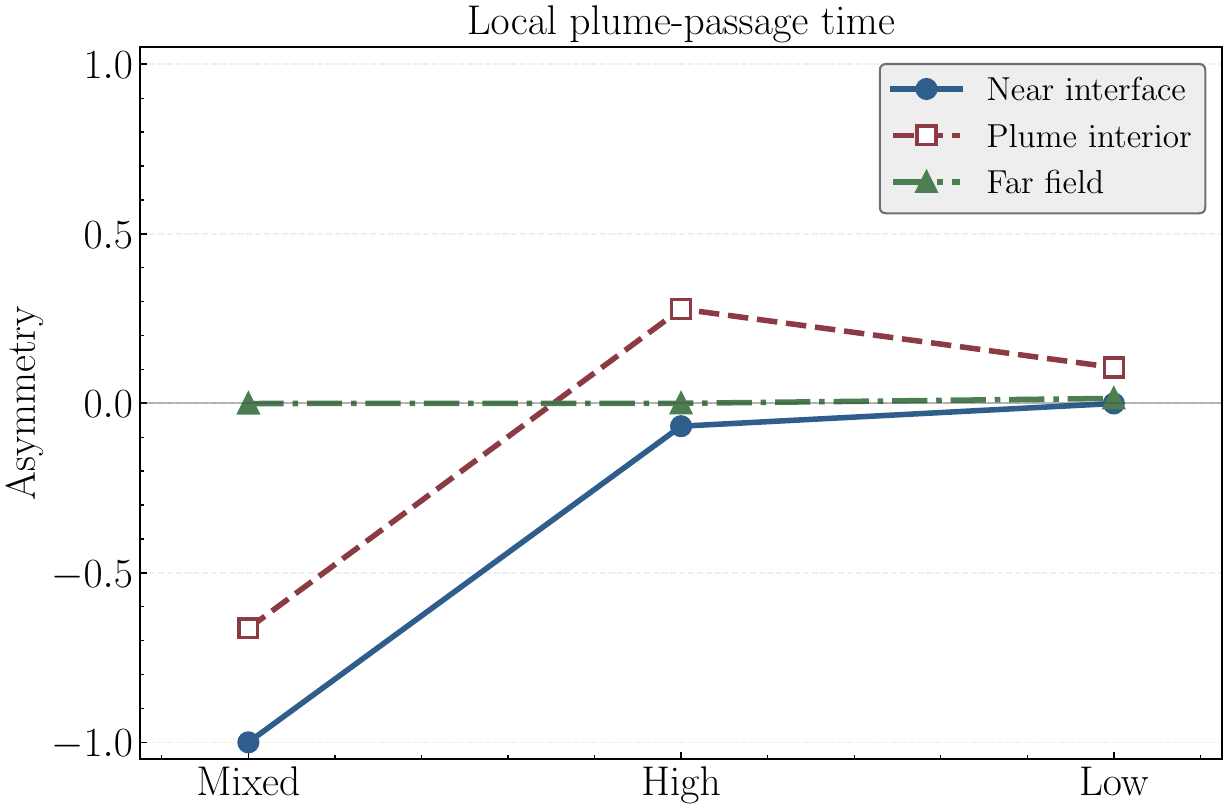}
    \caption{Passage duration}
    \label{fig:probe_event_time_fraction_asymmetry}
  \end{subfigure}\hfill
  \begin{subfigure}{0.32\linewidth}
    \centering
    \includegraphics[width=\linewidth]{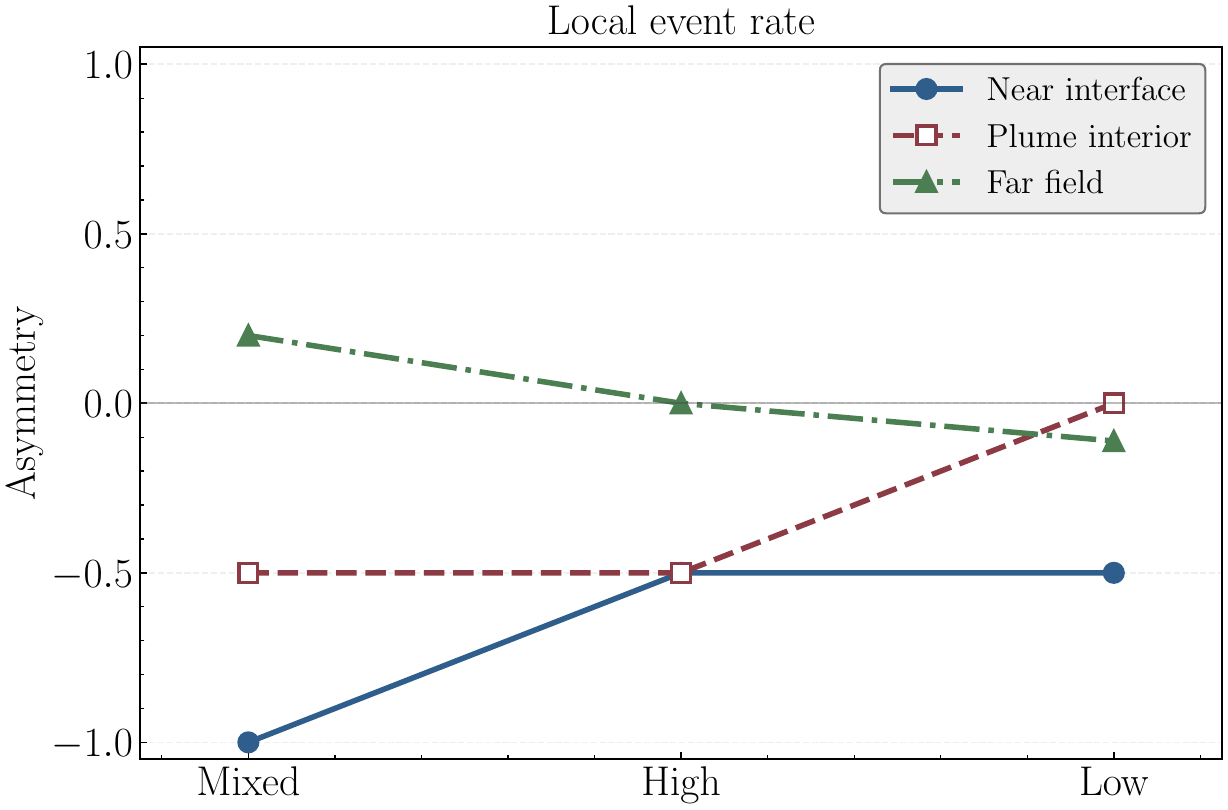}
    \caption{Event rate}
    \label{fig:probe_event_rate_asymmetry}
  \end{subfigure}\hfill
  \begin{subfigure}{0.32\linewidth}
    \centering
    \includegraphics[width=\linewidth]{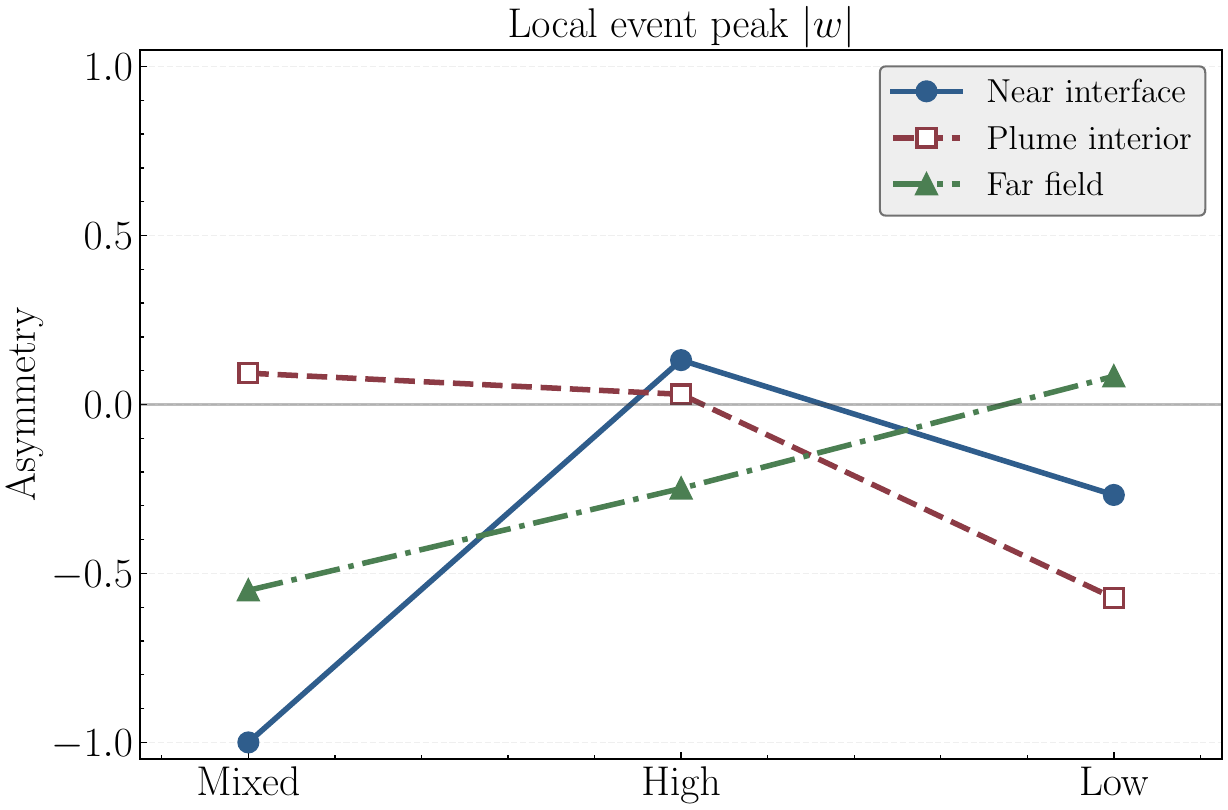}
    \caption{Peak amplitude}
    \label{fig:probe_event_peak_asymmetry}
  \end{subfigure}
  \caption{Local plume-passage asymmetry at fixed probes. The mixed route
  develops the largest lower-biased post-transition plume-passage timing near
  the interface and in the plume interior, while the endpoint spectra remain
  closer to local parity. Negative values denote lower-biased activity.}
  \label{fig:probe_asymmetry_local}
\end{figure}

\begin{figure}[tbp]
  \centering
  \begin{subfigure}{0.48\linewidth}
    \centering
    \includegraphics[width=\linewidth]{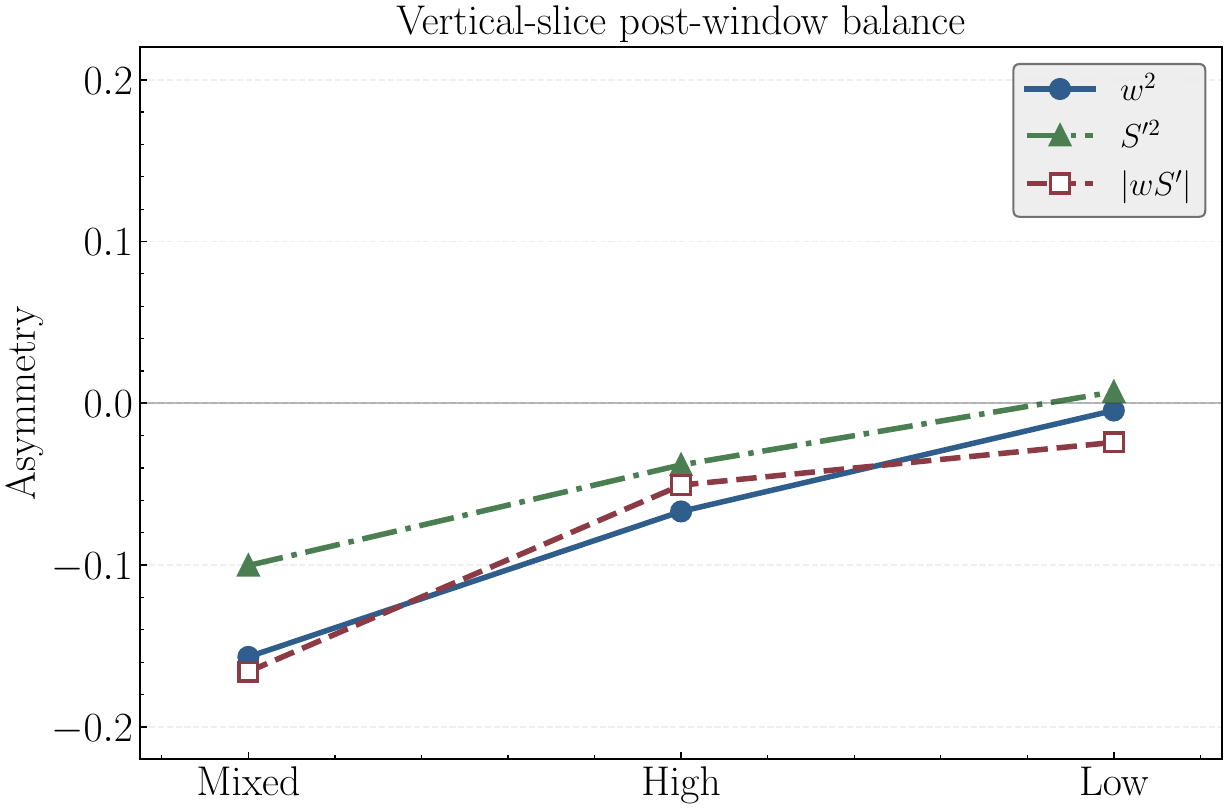}
    \caption{Vertical-slice asymmetry}
    \label{fig:field_slice_post_window_asymmetry}
  \end{subfigure}\hfill
  \begin{subfigure}{0.48\linewidth}
    \centering
    \includegraphics[width=\linewidth]{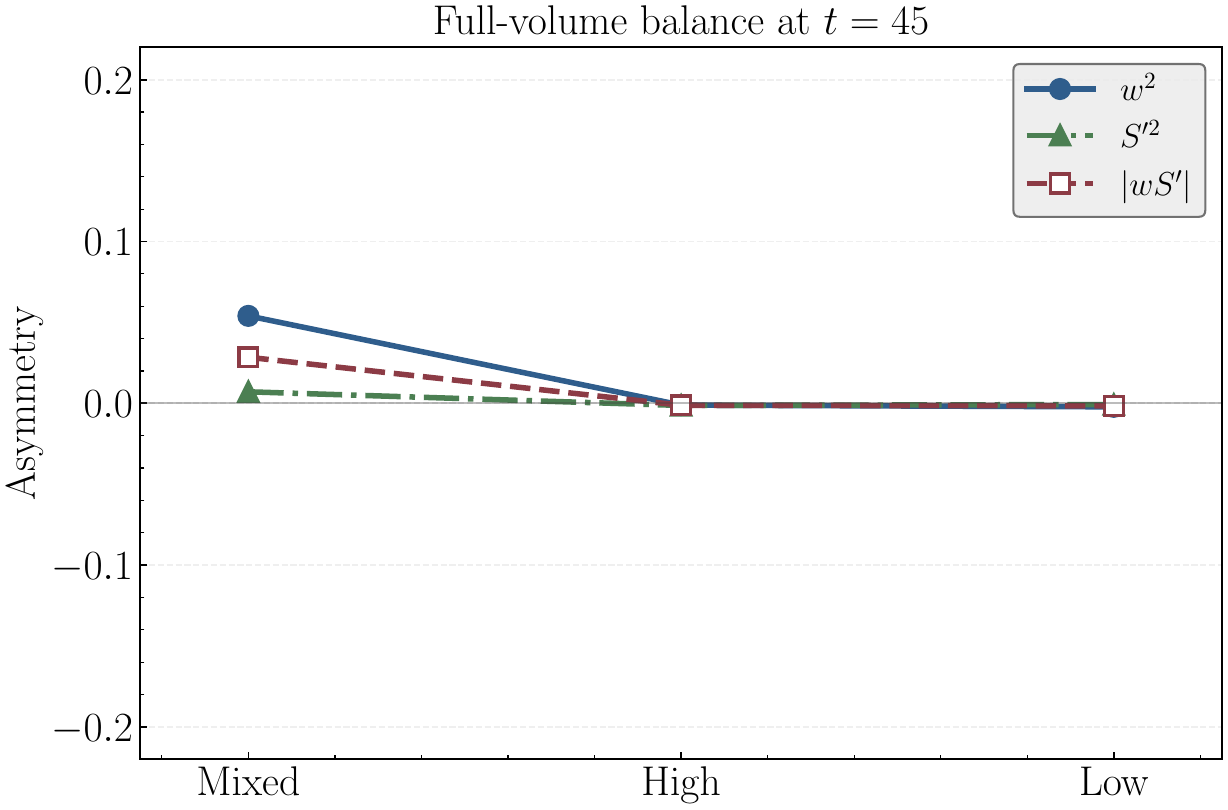}
    \caption{Full-volume asymmetry}
    \label{fig:volume_t45_asymmetry}
  \end{subfigure}
  \caption{Upper/lower balance from field measures. Vertical slices reveal
  local plume-passage bias, but the full-volume measure remains close to parity
  for all three spectra, showing that route selection does not require a
  domain-integrated upper/lower imbalance.}
  \label{fig:field_volume_asymmetry}
\end{figure}

Probe and upper/lower asymmetry measures (\cref{fig:probe_asymmetry_local,fig:field_volume_asymmetry}) place a second constraint on the route picture. They ask whether route selection produces a globally one-sided plume forest or instead produces local, intermittent plume-passage events within a nearly balanced volume. We use the asymmetry convention \((Q_{\mathrm{upper}}-Q_{\mathrm{lower}})/ (Q_{\mathrm{upper}}+Q_{\mathrm{lower}})\), so negative values are lower-biased. The passage-duration, event-rate, and peak-amplitude probe panels in
\cref{fig:probe_event_time_fraction_asymmetry,fig:probe_event_rate_asymmetry,fig:probe_event_peak_asymmetry}
show that local mixed-case probe event-time asymmetries can become large after the transition window. The near-interface event-time asymmetry is \(-1.000\), and the plume-interior event-time asymmetry is \(-0.664\). High-annulus and low-mode near-interface event-time asymmetries are much smaller, \(-0.067\) and approximately zero, respectively. The vertical-slice and full-volume panels in
\cref{fig:field_slice_post_window_asymmetry,fig:volume_t45_asymmetry} show
why that local bias should not be interpreted as a domain-wide imbalance: vertical-slice \(w^2\) asymmetry is lower-biased in the mixed case, with a post-window value of \(-0.157\), while the full-volume response remains much closer to parity.

\Cref{fig:field_volume_asymmetry} shows that full-volume interior balance is
much closer to parity than the local probes and slices. Local plume-passage asymmetry therefore coexists with a nearly balanced volume response, so the route-selection claims rely on field, spectral, envelope, and volume measures rather than isolated probe events. At \(t=45\), the mixed full-volume \(w^2\) asymmetry is \(0.054\), while the high-annulus and low-mode values are \(-0.001\) and \(-0.002\). The symmetric far-field treatment therefore does not force a one-sided forest; the probe records instead capture local intermittency and side-dependent passage timing.

\subsection{Transport strength and scalar spectral richness}
\label{sec:results_transport_spectra}

\begin{figure}[tbp]
  \centering
  \begin{subfigure}{0.48\linewidth}
    \centering
    \includegraphics[width=\linewidth]{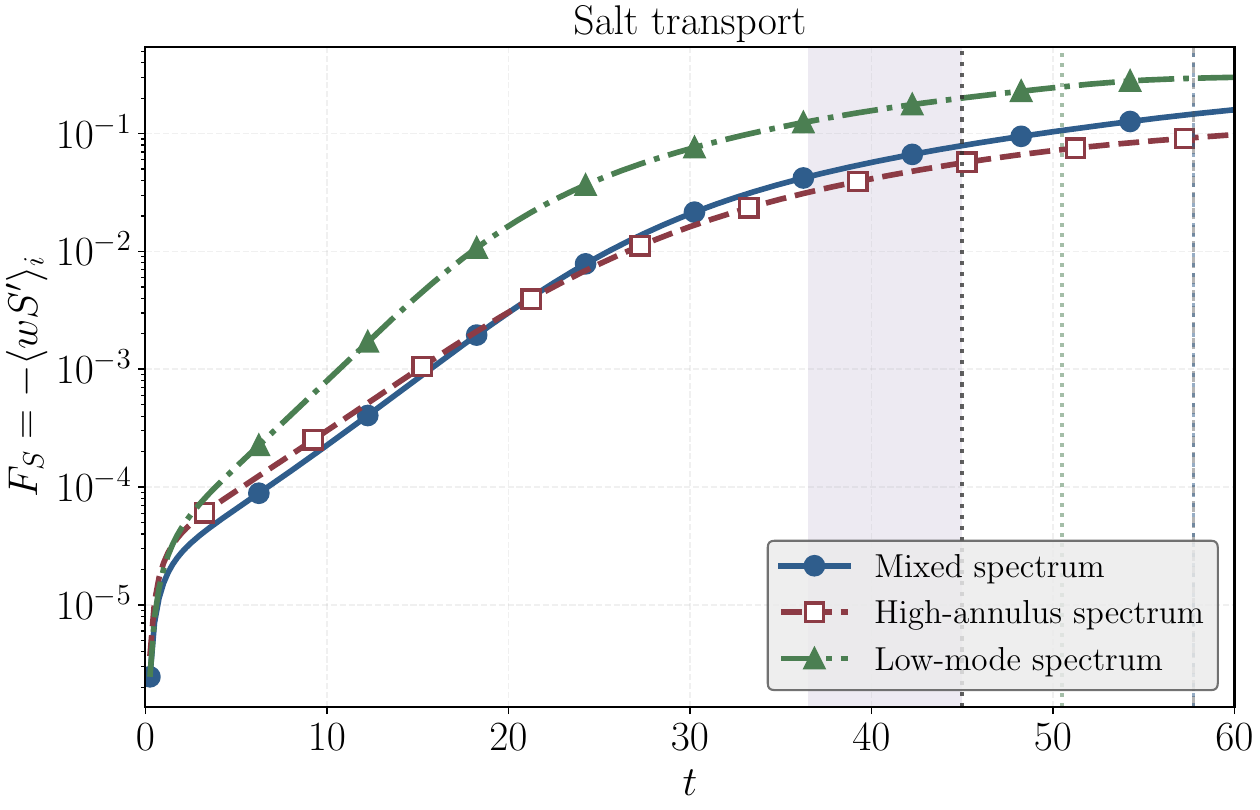}
    \caption{Salt transport}
    \label{fig:three_case_salt_transport}
  \end{subfigure}\hfill
  \begin{subfigure}{0.48\linewidth}
    \centering
    \includegraphics[width=\linewidth]{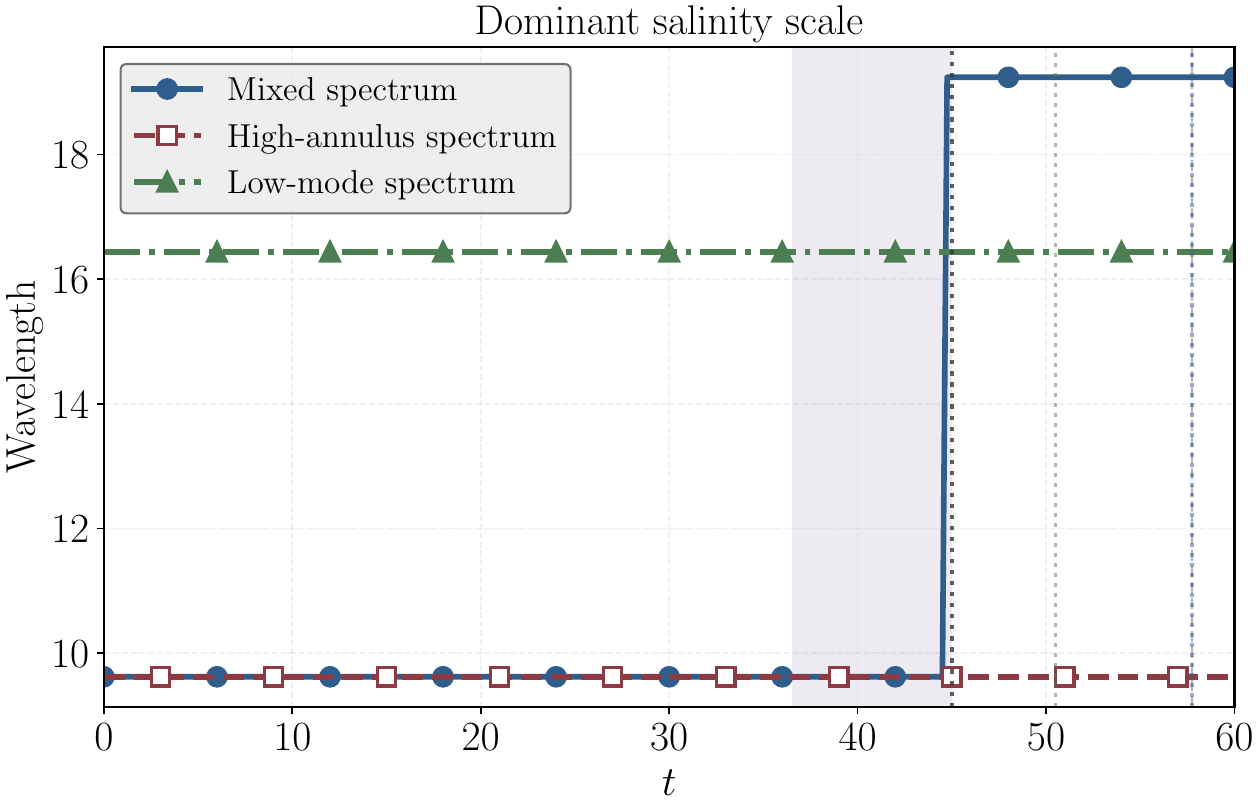}
    \caption{Dominant salinity wavelength}
    \label{fig:three_case_salinity_wavelength}
  \end{subfigure}
  \vspace{0.6em}
  \begin{subfigure}{0.48\linewidth}
    \centering
    \includegraphics[width=\linewidth]{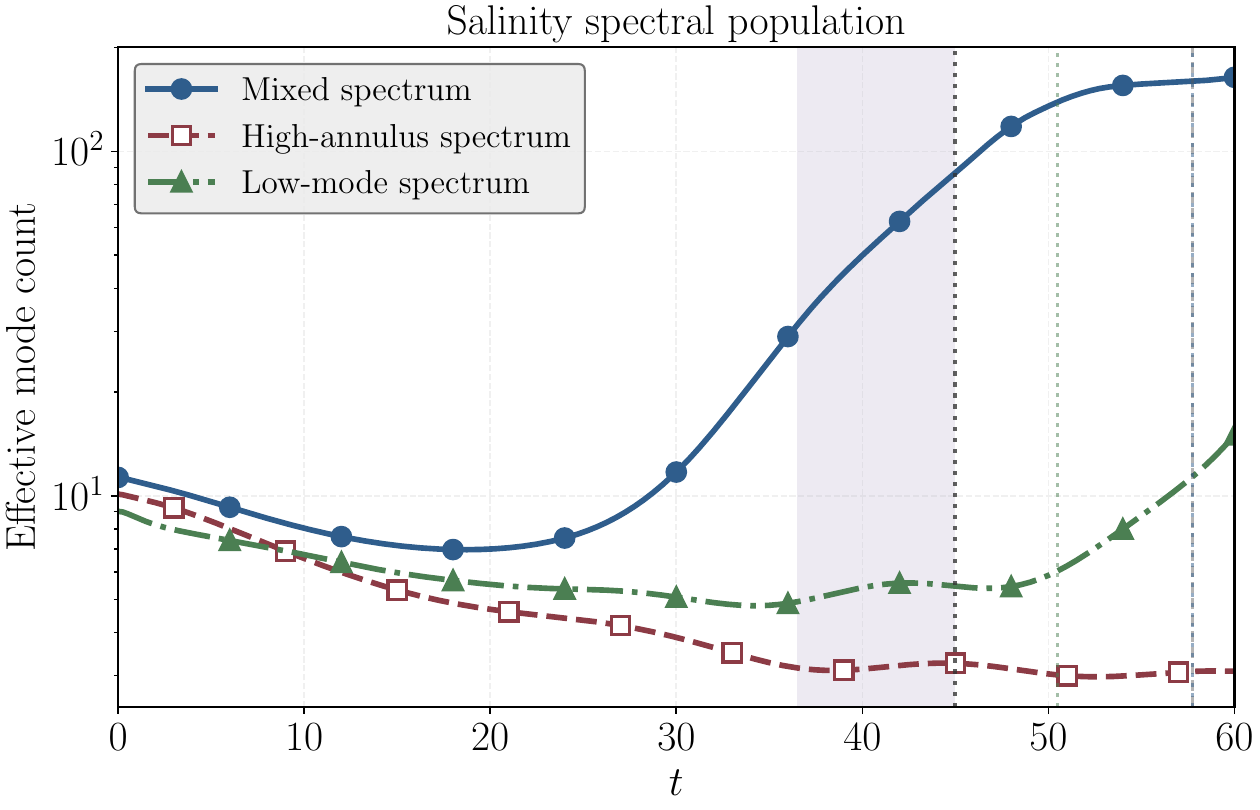}
    \caption{Salinity spectral population}
    \label{fig:three_case_salinity_effective_modes}
  \end{subfigure}\hfill
  \begin{subfigure}{0.48\linewidth}
    \centering
    \includegraphics[width=\linewidth]{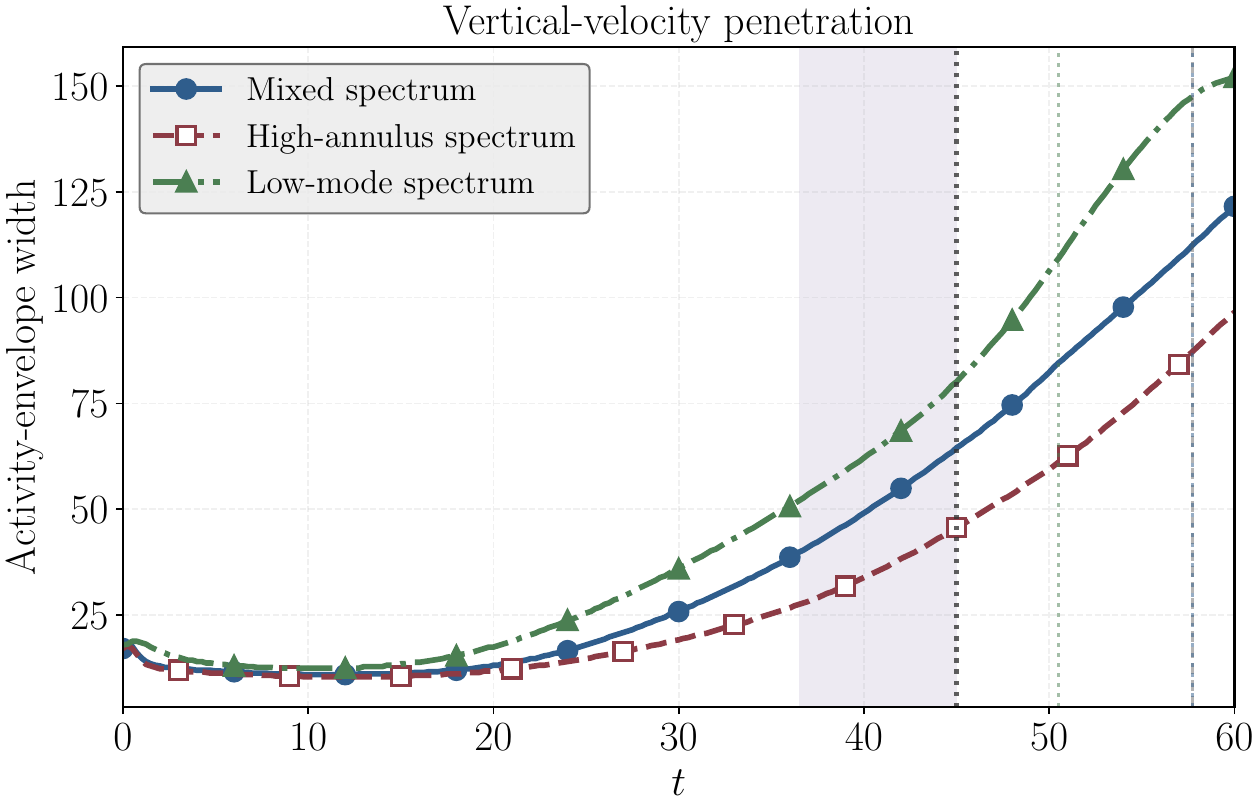}
    \caption{Vertical-velocity activity width}
    \label{fig:three_case_w_activity_width}
  \end{subfigure}
  \caption{Mechanism summary for the three imposed spectra. Low-mode forcing
  gives the largest interior transport and vertical reach, high-annulus forcing
  remains compact and weakly transporting, and mixed forcing is
  transport-intermediate while developing the broadest scalar spectral
  population.}
  \label{fig:three_case_mechanism_summary}
\end{figure}

The transport and spectral measures summarized in
\cref{fig:three_case_mechanism_summary,tab:mechanism_clean_t45} show that route selection does not
follow a monotonic dependence on imposed wavelength. At the interior comparison time, the low-mode case has the largest down-gradient salinity transport, kinetic energy, \(w_{\mathrm{rms}}\), and vertical-velocity activity width, whereas the high-annulus case is the weakest transport endpoint. In this finite-depth setting, imposing shorter horizontal scales does not intensify the interior-comparison transport. The exact values in \cref{tab:mechanism_clean_t45} give the scale of this separation: at \(t=45\), low-mode forcing has \(F_S=0.20068\), compared with \(0.07921\) for mixed forcing and \(0.05642\) for high-annulus forcing.

\begin{table}[tbp]
\centering
\caption{Interior-comparison transport, scale, spectral-population, and activity-width metrics at \(t=45\). Low-mode forcing maximizes interior transport and active width, whereas mixed forcing has the largest salinity spectral population.}
\label{tab:mechanism_clean_t45}
\begin{tabular}{lccc}
\toprule
Metric & Mixed & High-annulus & Low-mode \\
\midrule
\(F_S=-\langle wS'\rangle_i\) & 0.07921 & 0.05642 & 0.2007 \\
Salinity dominant wavelength & 19.24 & 9.619 & 16.44 \\
Salinity effective modes & 86.66 & 3.26 & 5.459 \\
\(w\) activity width & 64.55 & 45.71 & 80.13 \\
\bottomrule
\end{tabular}
\end{table}

\Cref{fig:three_case_mechanism_summary,tab:mechanism_clean_t45} show that the mixed case is
transport-intermediate at \(t=45\). The salt-transport and activity-width panels
\cref{fig:three_case_salt_transport,fig:three_case_w_activity_width} place
that intermediate transport beside the finite-depth reach. Its distinguishing feature is the combination of broad organizing scale and strong scalar spectral population. The dominant-wavelength and effective-mode panels
\cref{fig:three_case_salinity_wavelength,fig:three_case_salinity_effective_modes}
show that the mixed case has the broadest dominant salinity wavelength and a salinity effective mode count far larger than either single-band endpoint. Its \(t=45\) salinity effective mode count is \(86.66\), compared with \(3.26\) for high-annulus forcing and \(5.46\) for low-mode forcing; by \(t=57.75\), the mixed value reaches \(159.88\). Thus the mixed scalar field broadens spectrally even though it is not the maximum-flux route.

The effective-mode result in \cref{fig:three_case_mechanism_summary} is especially important. Low-mode forcing gives the largest flux while retaining a relatively narrow active modal population; high-annulus forcing remains spectrally narrow because it stays branch-locked. The mixed scalar field instead develops a broad spectral population while its dominant scale reorganizes.

\begin{figure}[tbp]
  \centering
  \includegraphics[width=\linewidth]{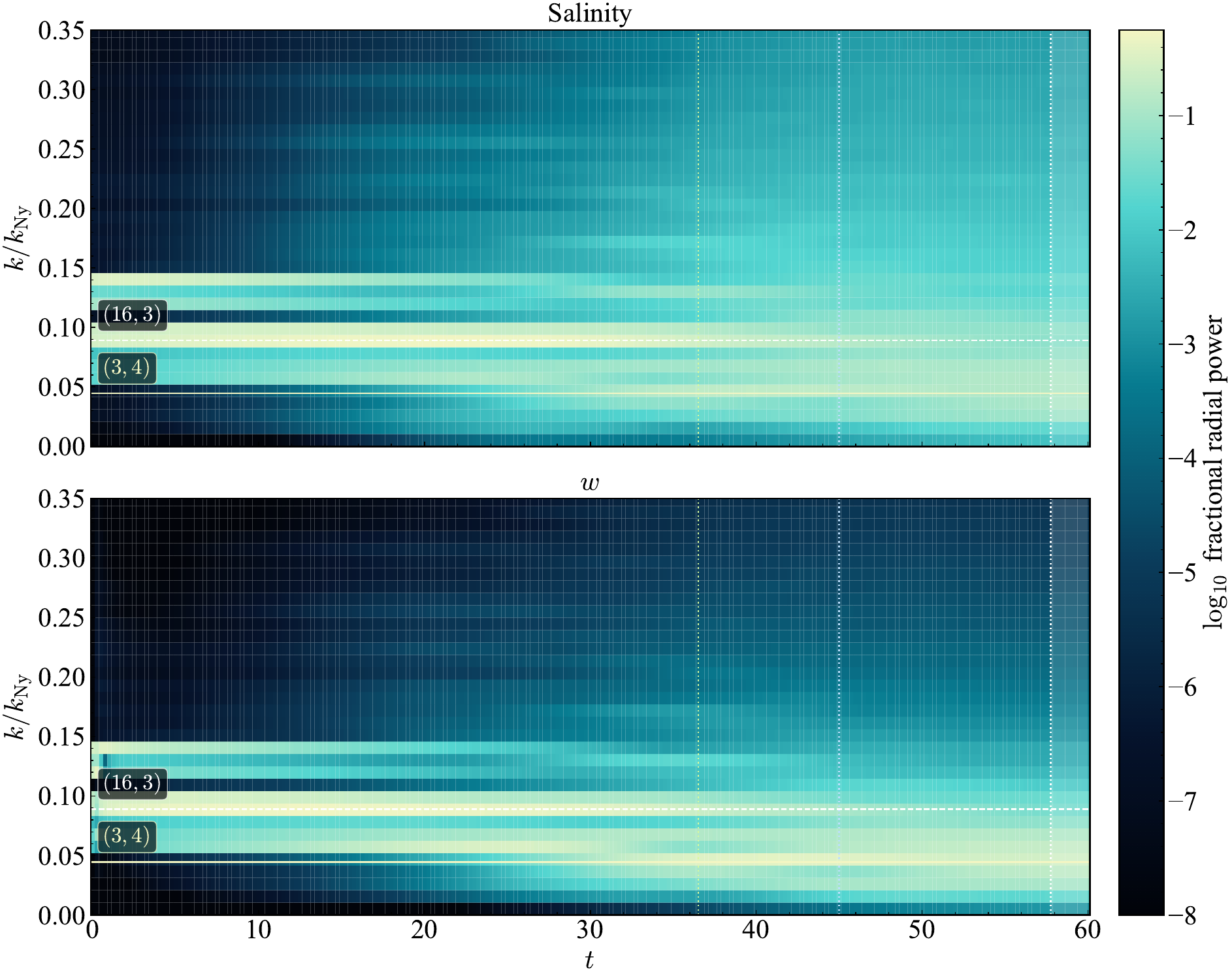}
  \caption{Mixed-spectrum time-wavenumber redistribution. The salinity
  dominant radial bin moves toward the broad route while the populated scalar
  wavenumber range grows. The velocity field shows a simpler coarsening path,
  whereas salinity coarsens and broadens at the same time.}
  \label{fig:time_wavenumber_main}
\end{figure}

\Cref{fig:time_wavenumber_main} shows why the mixed case is more than a
coarsening event. The dominant salinity wavelength increases from the imposed short-scale value to the broad value, while the scalar field fills a larger set of planform modes. The organizing wavelength becomes broad, yet the salinity field is more spectrally populated than either single-band response. The high-annulus response keeps short-scale branch identity without broad population growth. The low-mode response transports strongly without the same scalar spectral richness. Mixed forcing occupies the distinct middle route, combining broad dominant organization with scalar spectral broadening.

\begin{figure}[tbp]
  \centering
  \includegraphics[width=\linewidth]{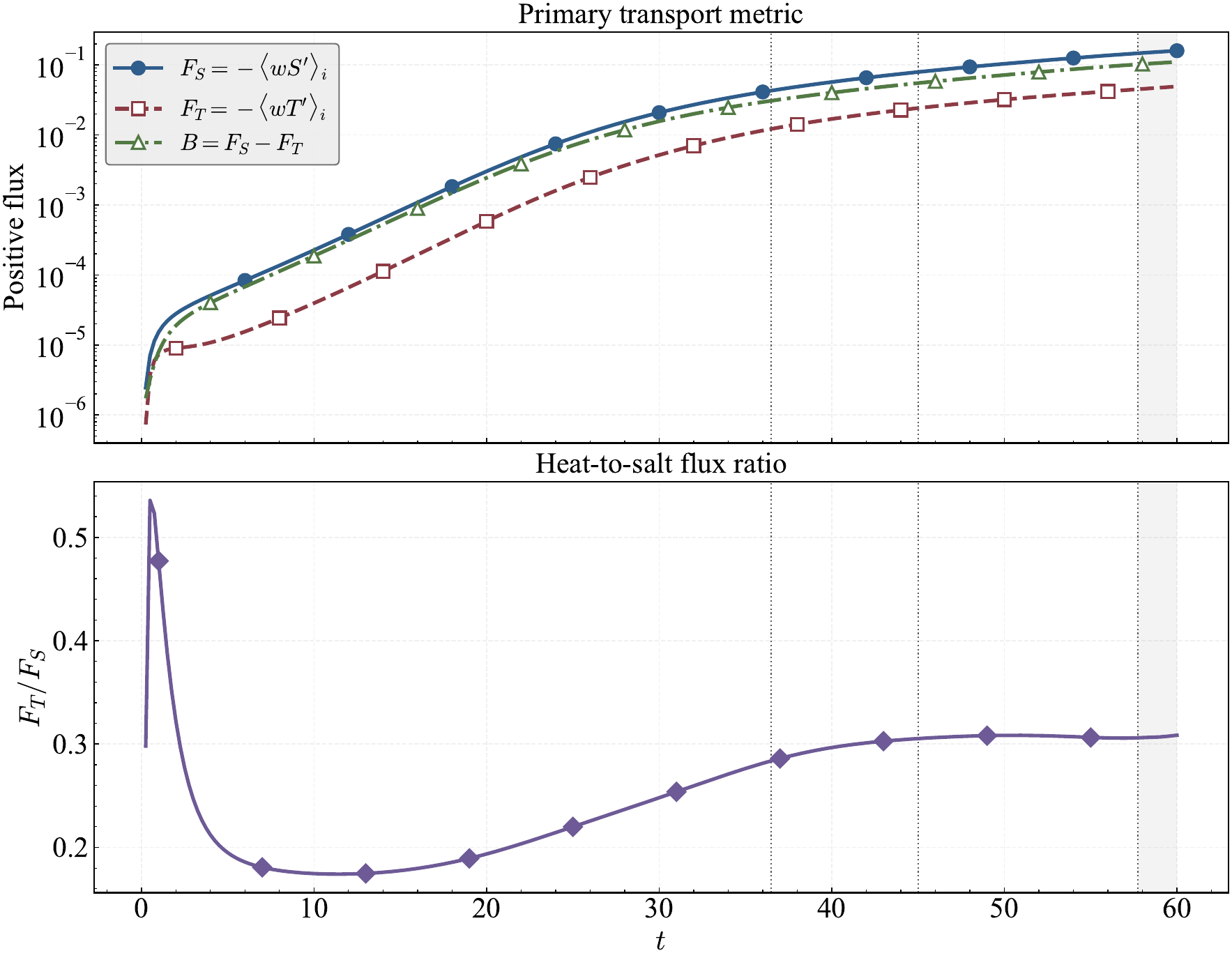}
  \caption{Primary salinity-transport history for the three imposed spectra.
  Low-mode forcing is the interior-comparison transport maximum, high-annulus forcing
  remains weakly transporting, and mixed forcing is transport-intermediate
  while retaining the distinct velocity-led spectral route.}
  \label{fig:transport_primary_flux}
\end{figure}

\begin{figure}[tbp]
  \centering
  \includegraphics[width=\linewidth]{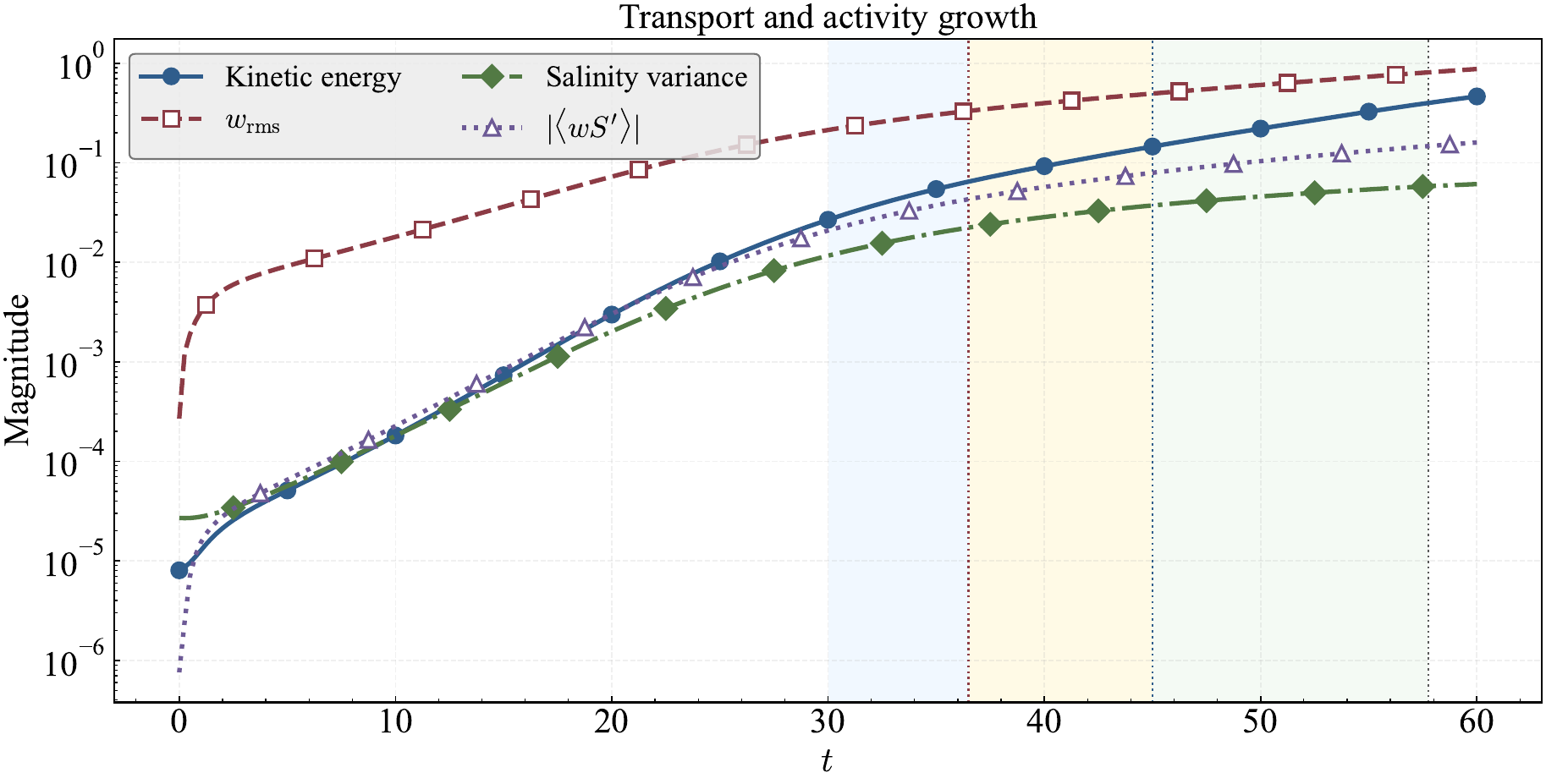}
  \caption{Mixed-spectrum transport-growth windows. The down-gradient salinity
  flux, kinetic energy, \(w_{\rm rms}\), and salinity variance grow through
  the modal reorganization interval, while the fitted growth rates decrease from
  the pre-transition to post-transition windows.}
  \label{fig:transport_transition_slopes_main}
\end{figure}

The transport histories in \cref{fig:transport_primary_flux,fig:transport_transition_slopes_main} fix the main claim. The low-mode spectrum is the interior-comparison transport maximum among the three spectra, while the mixed spectrum is the clearest example of nonlinear route selection. The physics is not a single-wavelength ranking. The initial spectrum controls whether the system remains locked, immediately selects a broad branch, or undergoes a delayed velocity-led handoff with scalar spectral broadening.

\Cref{fig:transport_transition_slopes_main} shows that the mixed
transport history does not contain a sudden transition-triggered burst. Instead, modal reorganization occurs inside an already amplifying finite-depth plume-growth phase. The fitted growth factors link planform reorganization to transport-relevant growth without implying that the modal crossing is a singular event or a universal transition time. In the baseline mixed case, \(F_S\) grows by a factor of \(1.842\) between the velocity crossing and salinity crossing window \((36.5 \le t \le 45)\), while kinetic energy grows by \(2.251\) and \(w_{\rm rms}\) by \(1.494\).

\subsection{Active-interface geometry and scalar interleaving}
\label{sec:results_interface_geometry}

The interface measures show that displacement, scalar-interface thickness, and scalar spectral population respond differently to the imposed spectrum. This distinction matters because, once fingers interleave, a single zero-crossing surface no longer describes the active salinity structure. The temperature-zero surface tracks two-layer interface displacement, while salinity-gradient measures track the interleaved active scalar layer after fingering has developed.

\begin{figure}[tbp]
  \centering
  \begin{subfigure}{0.48\linewidth}
    \centering
    \includegraphics[width=\linewidth]{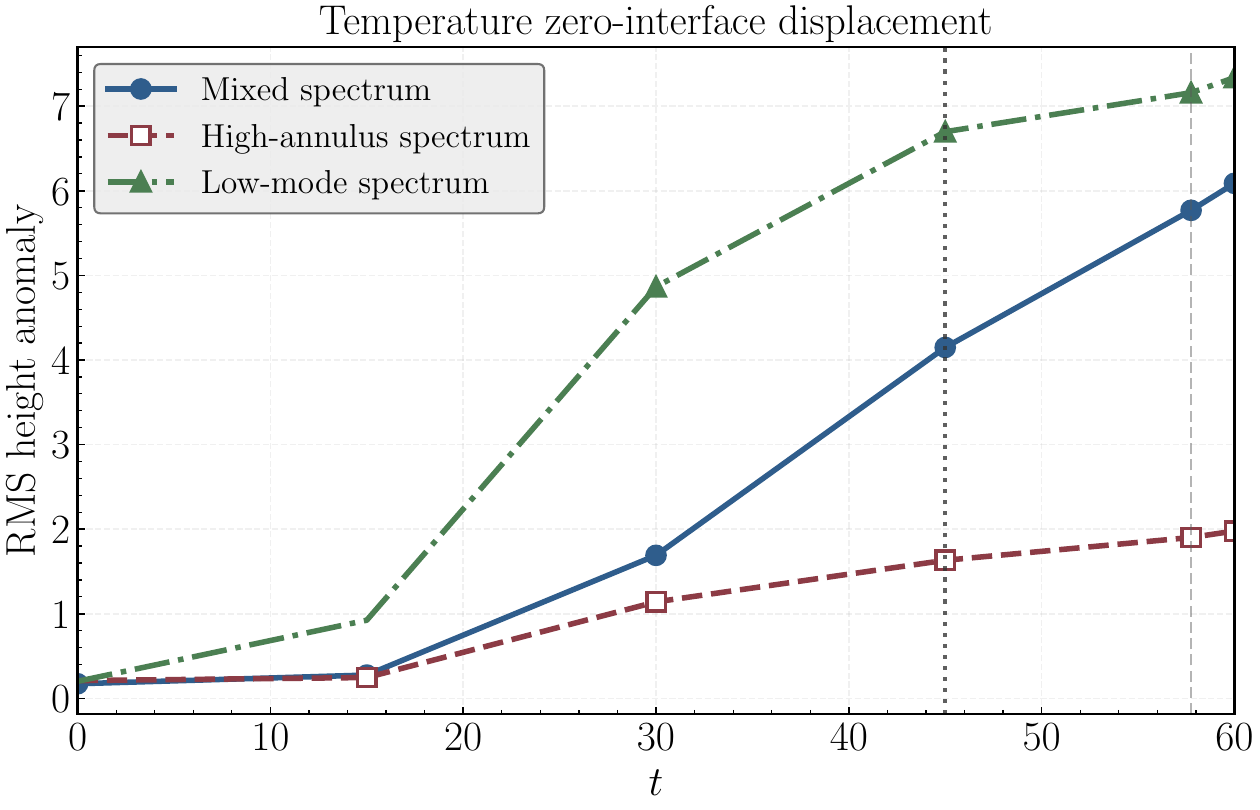}
    \caption{Temperature-zero height}
    \label{fig:interface_T_zero_rms_height}
  \end{subfigure}\hfill
  \begin{subfigure}{0.48\linewidth}
    \centering
    \includegraphics[width=\linewidth]{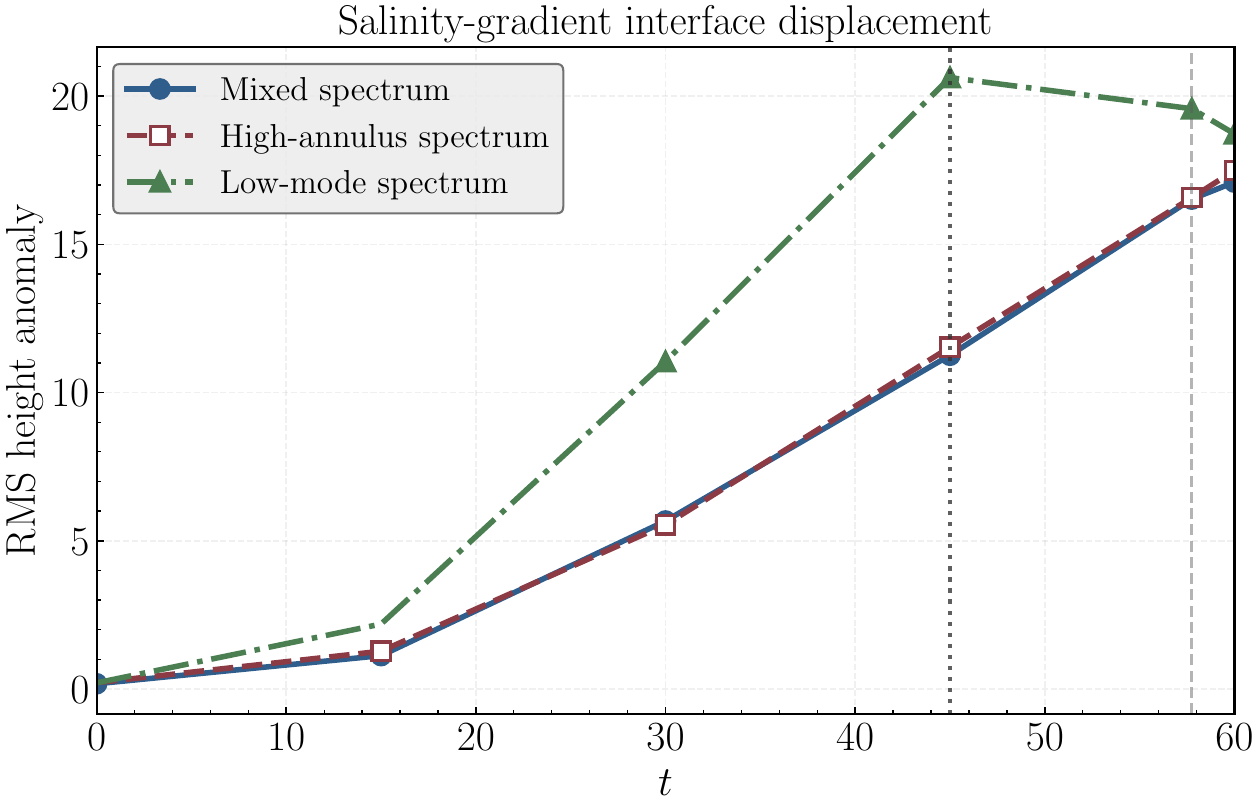}
    \caption{Salinity-gradient height}
    \label{fig:interface_S_gradient_rms_height}
  \end{subfigure}
  \vspace{0.6em}
  \begin{subfigure}{0.48\linewidth}
    \centering
    \includegraphics[width=\linewidth]{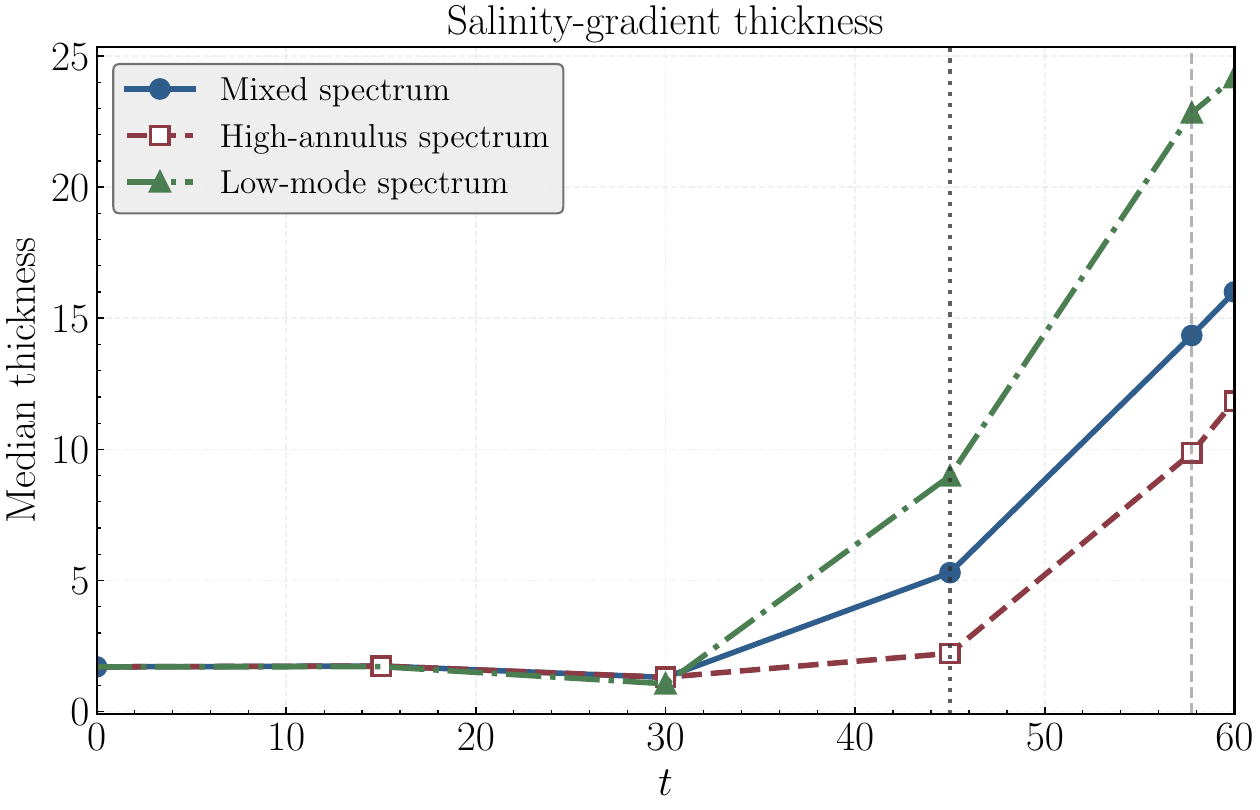}
    \caption{Salinity-gradient thickness}
    \label{fig:interface_S_gradient_thickness}
  \end{subfigure}\hfill
  \begin{subfigure}{0.48\linewidth}
    \centering
    \includegraphics[width=\linewidth]{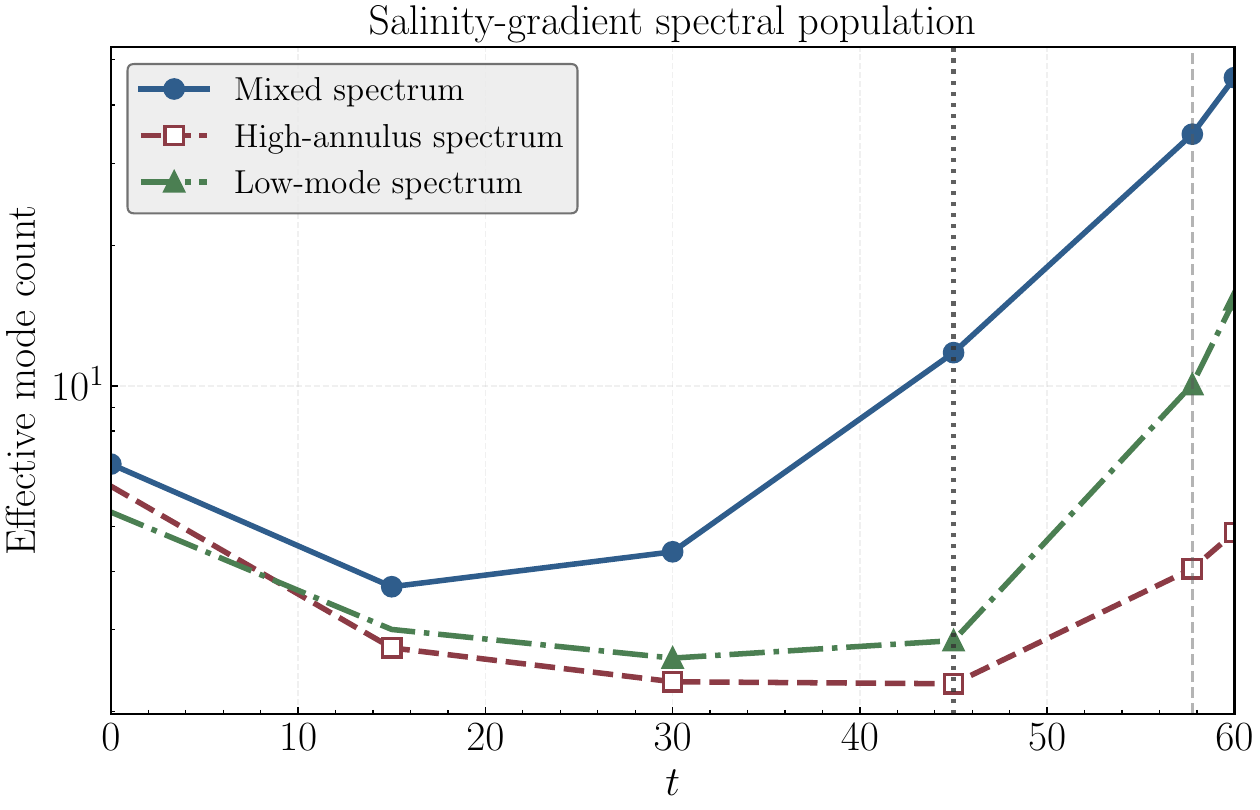}
    \caption{Salinity-gradient modes}
    \label{fig:interface_S_gradient_effective_modes}
  \end{subfigure}
  \caption{Interface geometry at the interior-comparison time. The
  temperature-zero surface measures cleaner two-layer displacement, whereas
  salinity-gradient metrics track the active interleaving layer. Low-mode
  forcing gives the largest displacement and thickness, while mixed forcing
  gives the richest active scalar interface.}
  \label{fig:three_case_interface_geometry}
\end{figure}

\begin{table}[tbp]
\centering
\caption{Interface-geometry measures at \(t=45\). The temperature-zero surface tracks cleaner displacement, while salinity-gradient metrics track the interleaved active scalar layer.}
\label{tab:interface_geometry_t45}
\begin{tabular}{lcccc}
\toprule
Case & \(T=0\) RMS \(h\) & \(S_z\) RMS \(h\) & \(S_z\) med. thick. & \(S_z\) modes \\
\midrule
Mixed spectrum & 4.148 & 11.26 & 5.298 & 11.78 \\
High-annulus spectrum & 1.632 & 11.53 & 2.221 & 2.297 \\
Low-mode spectrum & 6.698 & 20.62 & 8.989 & 2.843 \\
\bottomrule
\end{tabular}
\end{table}

\Cref{fig:three_case_interface_geometry,tab:interface_geometry_t45} show that,
at \(t=45\), the low-mode case has the largest temperature-zero displacement, salinity-gradient centroid height, and active-layer thickness. The high-annulus case is smaller and thinner, while the mixed case is intermediate in displacement and thickness. The low-mode temperature-zero RMS height is \(6.698\), compared with \(4.148\) for mixed forcing and \(1.632\) for high-annulus forcing. The corresponding median salinity-gradient thicknesses are \(8.989\), \(5.298\), and \(2.221\), respectively.

The salinity-gradient effective-mode panel
\cref{fig:interface_S_gradient_effective_modes} gives the complementary
ranking: mixed forcing produces the richest gradient-based scalar interface. Transport strength, active-layer thickening, and scalar spectral richness are therefore related but distinct responses to the imposed interfacial spectrum. The mixed salinity-gradient effective mode count is \(11.777\), larger than the high-annulus and low-mode values of \(2.297\) and \(2.843\).

\begin{figure}[tbp]
  \centering
  \includegraphics[width=\linewidth]{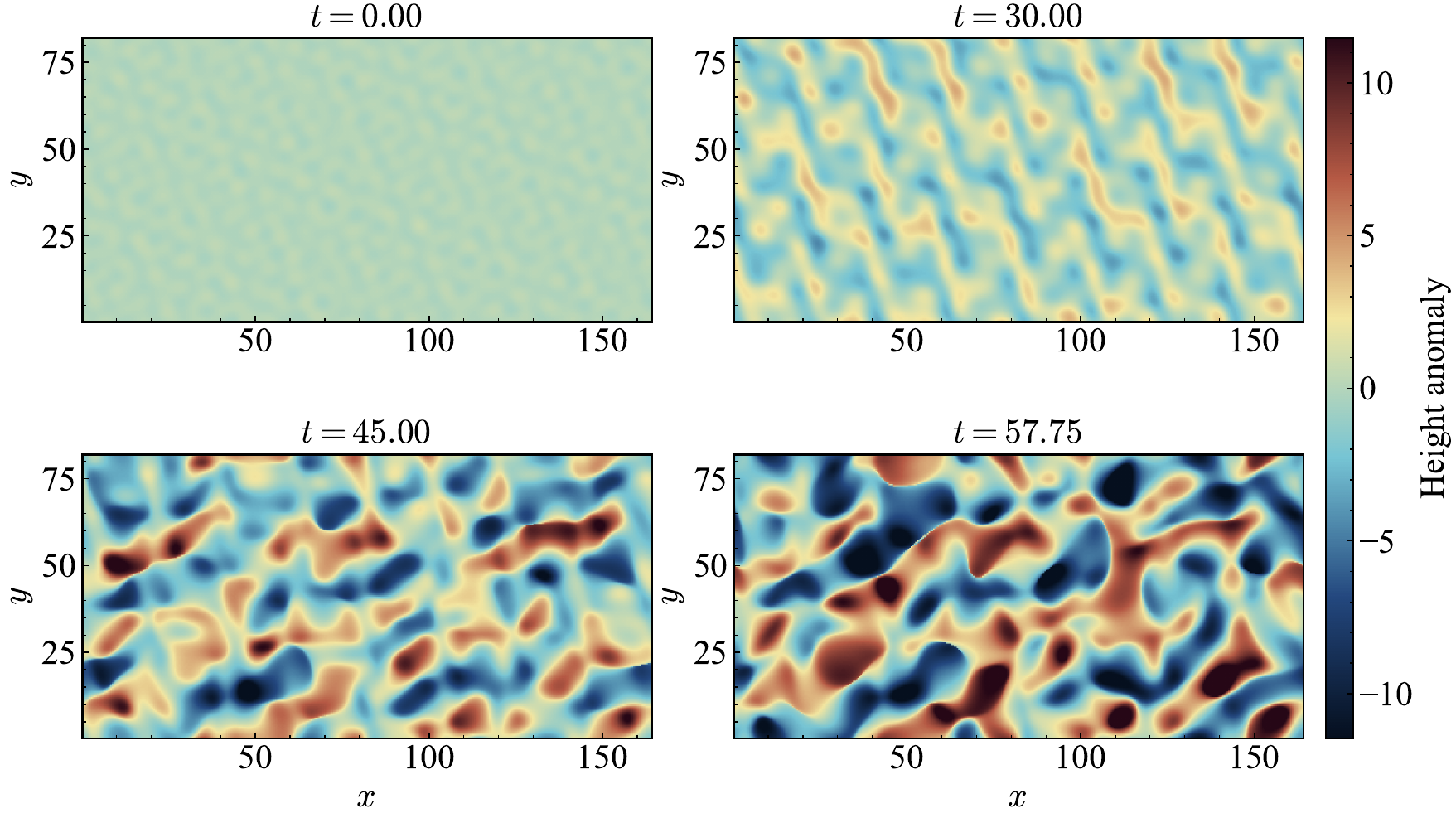}
  \includegraphics[width=\linewidth]{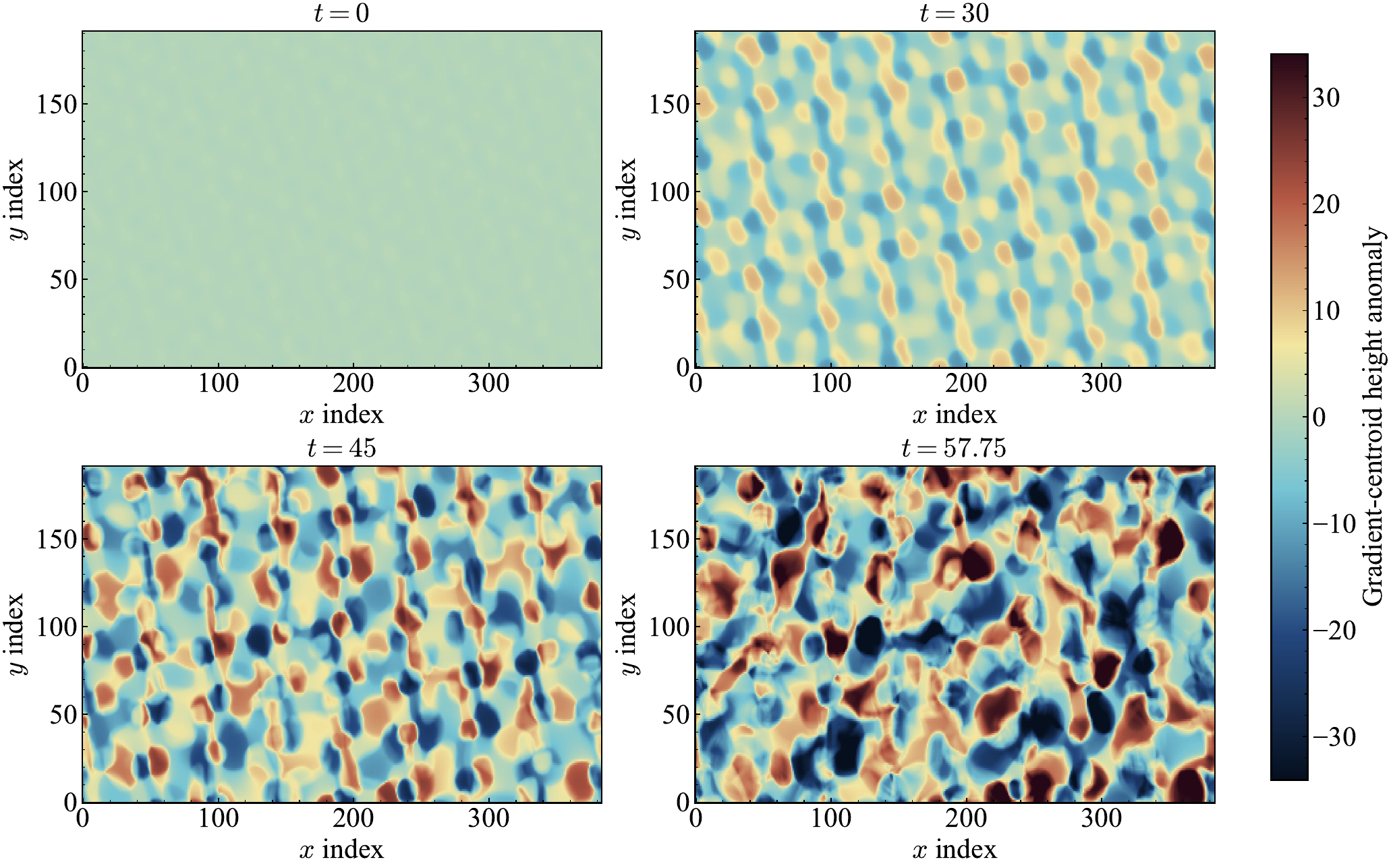}
  \caption{Spatial interface visualization through the interior-growth window.
  The temperature-zero maps show the displaced two-layer interface, while the
  salinity-gradient maps show the active interleaving layer. The spatial maps
  confirm the summary metrics: low-mode forcing produces the largest displaced
  and thickened interface, whereas mixed forcing produces a scalar-gradient
  interface with richer spatial structure.}
  \label{fig:interface_spatial_maps}
\end{figure}

The spatial maps in \cref{fig:interface_spatial_maps} reinforce the route-selection picture from a different perspective. Low-mode forcing creates a large displaced and thickened active layer, consistent with its strong transport and early finite-depth reach. High-annulus forcing produces a thinner and more compact active layer. Mixed forcing produces a scalar interface with spectrally richer active-gradient structure. At the interface level, the mixed route combines strong scalar interleaving and spectral population without maximum interior-comparison flux.

\subsection{Finite-depth reach and boundary approach}
\label{sec:results_penetration}

\begin{figure}[tbp]
  \centering
  \begin{subfigure}{0.48\linewidth}
    \centering
    \includegraphics[width=\linewidth]{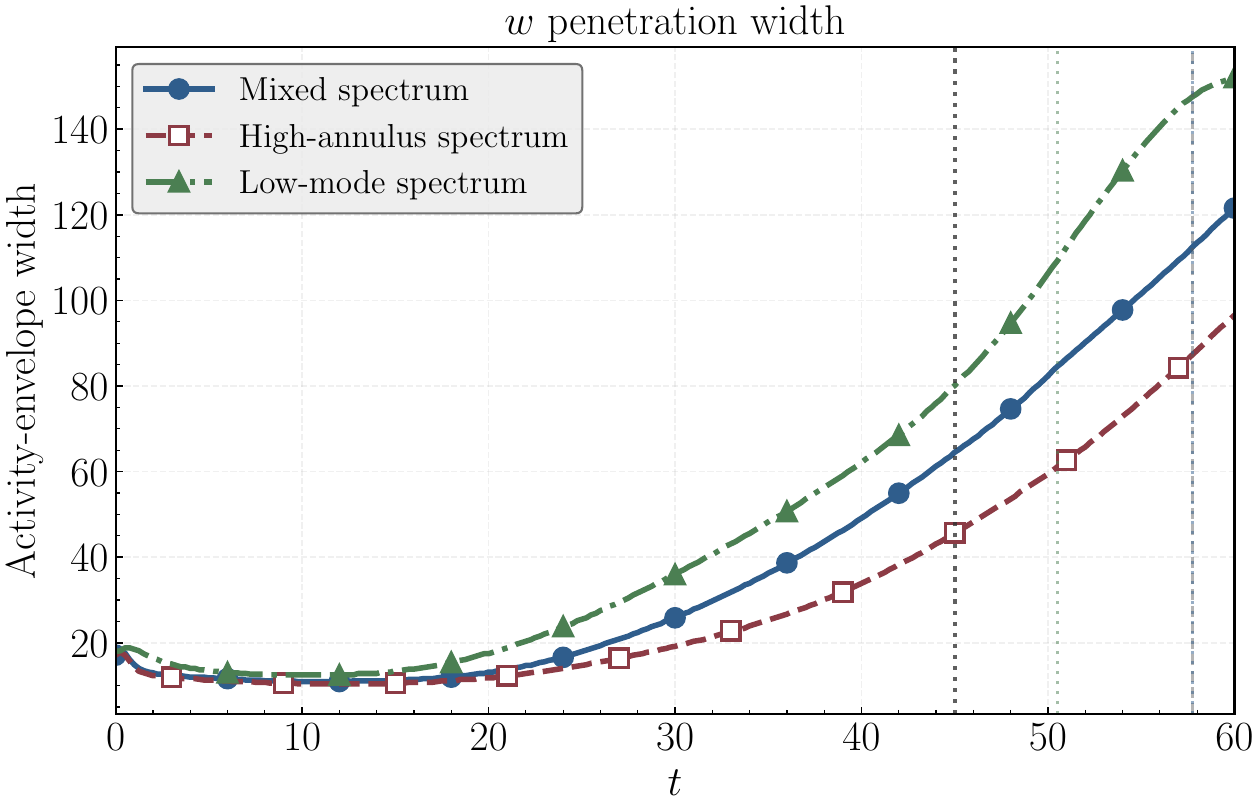}
    \caption{Vertical-velocity activity width}
    \label{fig:vertical_penetration_w_width}
  \end{subfigure}\hfill
  \begin{subfigure}{0.48\linewidth}
    \centering
    \includegraphics[width=\linewidth]{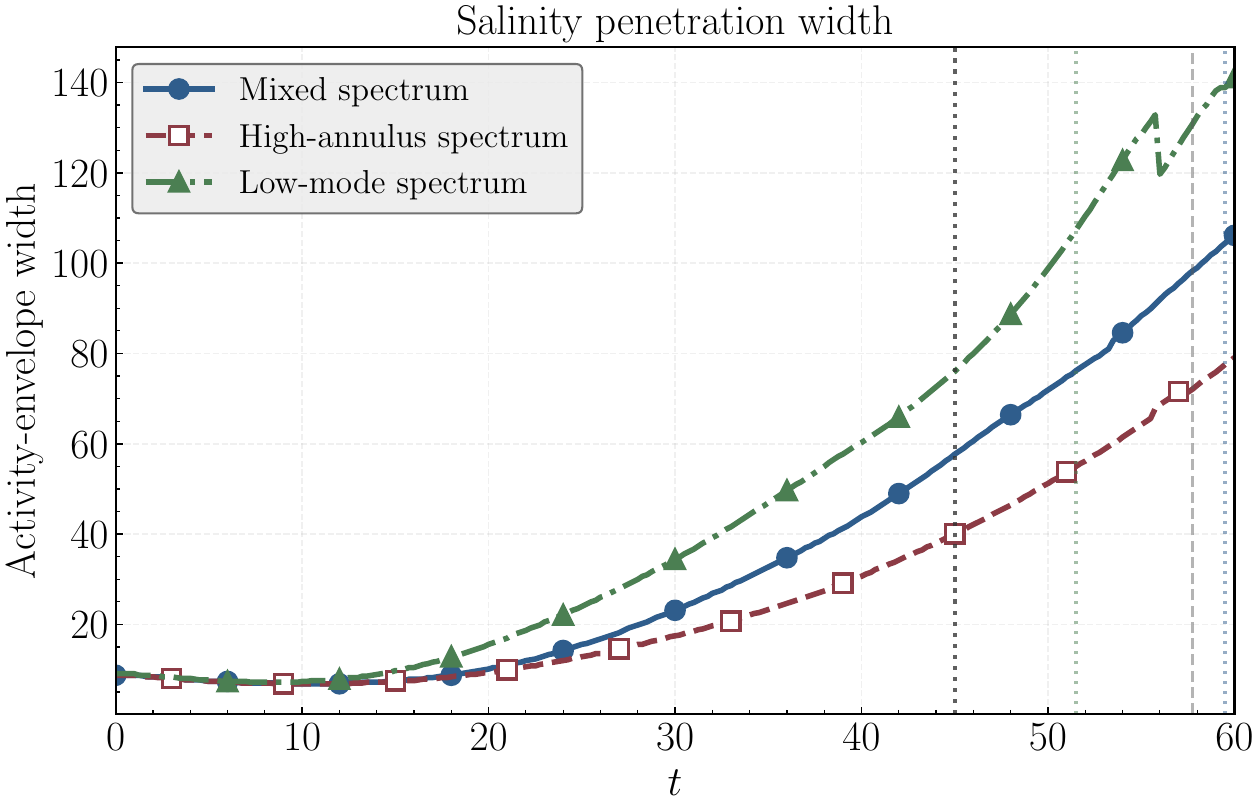}
    \caption{Salinity-activity width}
    \label{fig:vertical_penetration_S_width}
  \end{subfigure}
  \vspace{0.6em}
  \begin{subfigure}{0.48\linewidth}
    \centering
    \includegraphics[width=\linewidth]{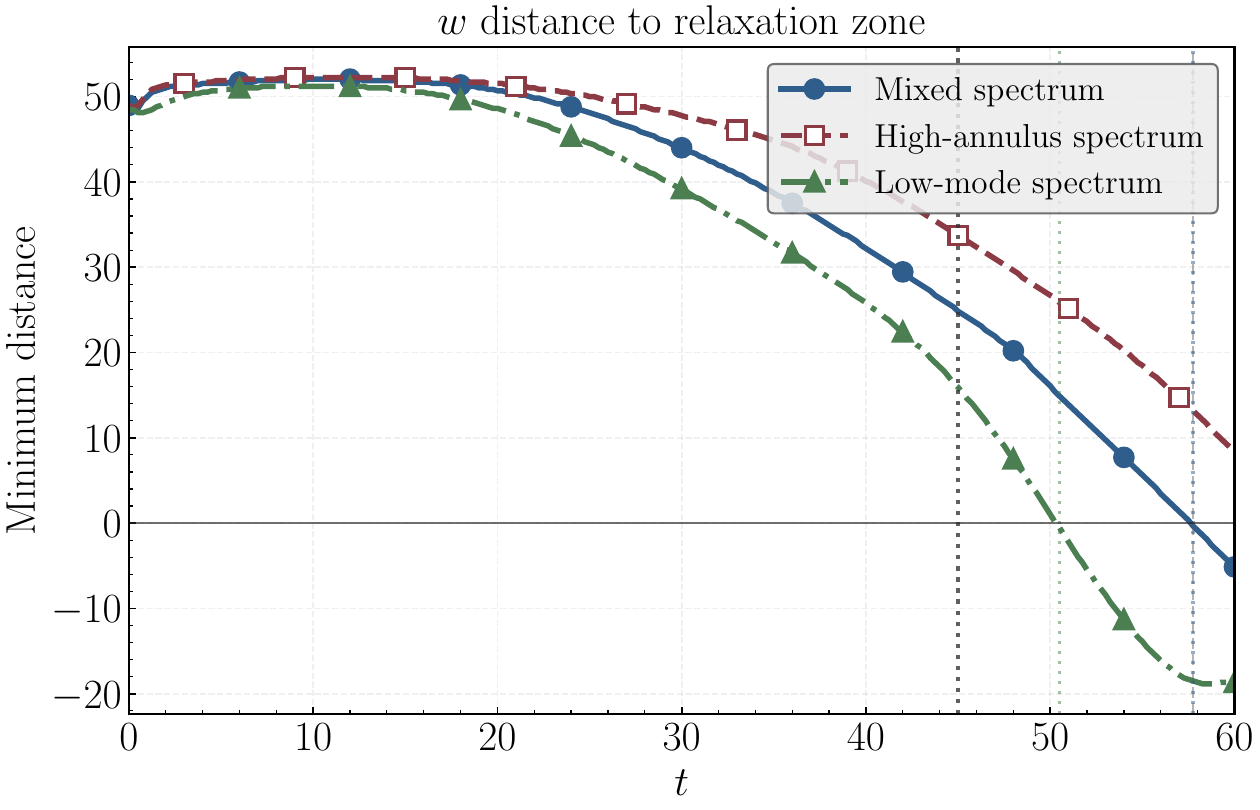}
    \caption{Vertical-velocity activity distance}
    \label{fig:vertical_penetration_w_relaxation_distance}
  \end{subfigure}\hfill
  \begin{subfigure}{0.48\linewidth}
    \centering
    \includegraphics[width=\linewidth]{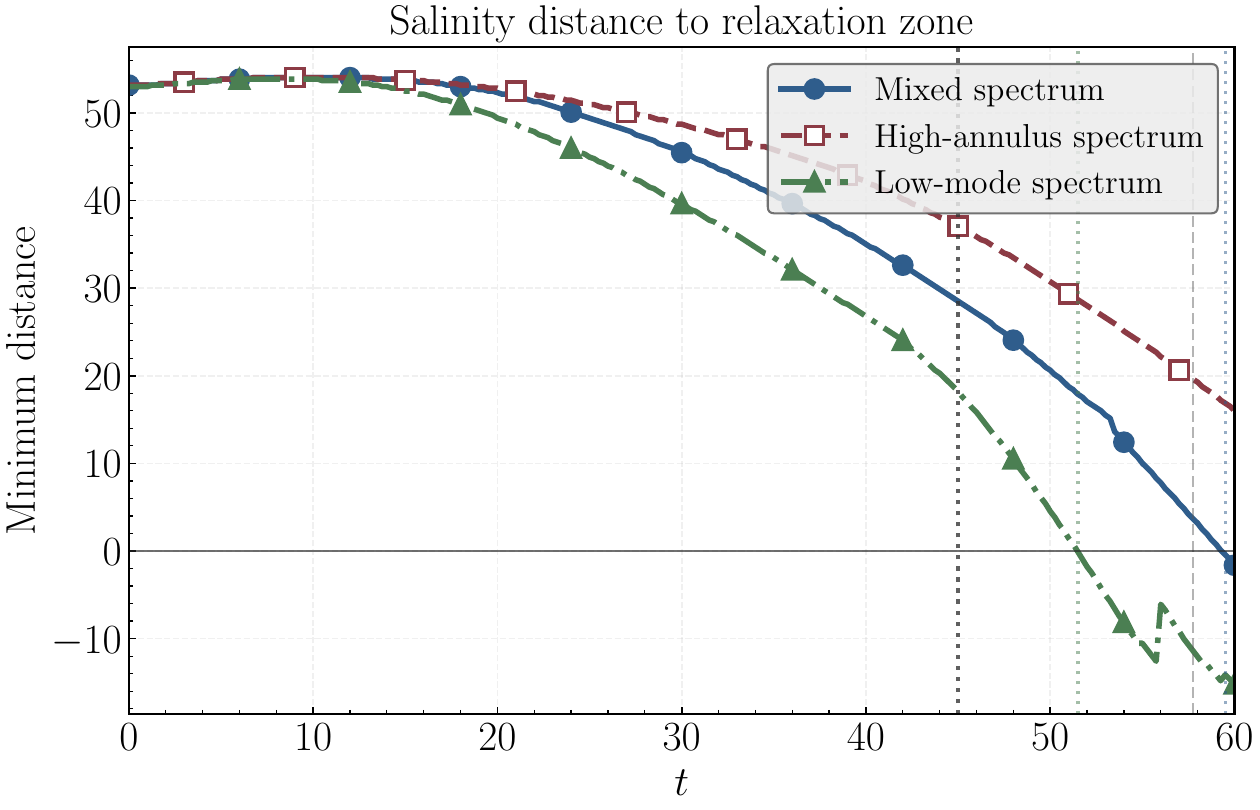}
    \caption{Salinity-activity distance}
    \label{fig:vertical_penetration_S_relaxation_distance}
  \end{subfigure}
  \caption{Finite-depth reach for the three spectra. Activity-envelope widths
  measure vertical spreading away from the original interface, while minimum
  distances to the inner edge of the relaxation zones define the standard
  boundary-region contact times used to separate interior comparison from later
  boundary approach.}
  \label{fig:three_case_vertical_penetration}
\end{figure}

\begin{figure}[tbp]
  \centering
  \includegraphics[width=\linewidth,height=0.62\textheight,keepaspectratio]{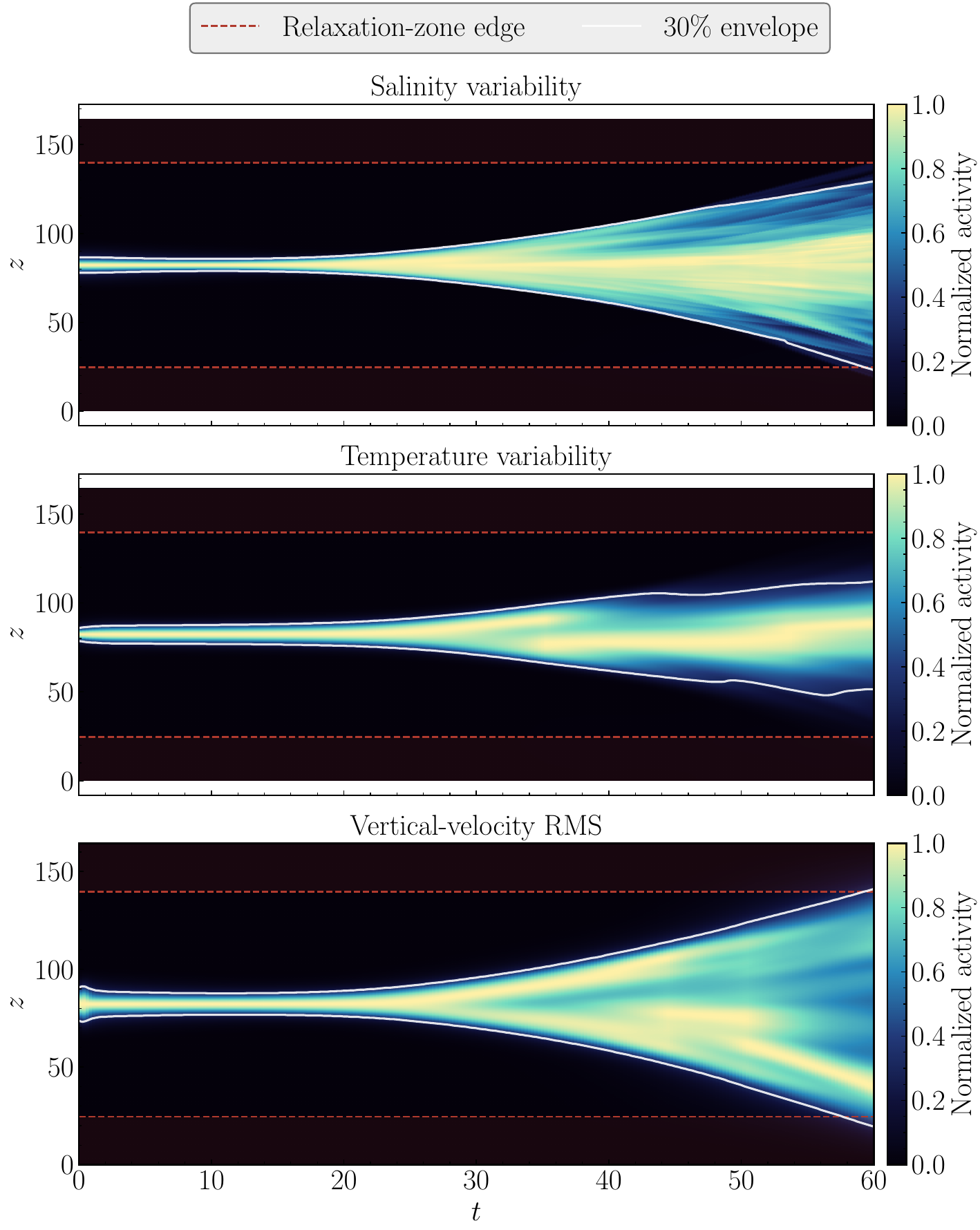}
  \caption{Time-dependent vertical activity envelopes for the mixed-spectrum
  route. The envelopes show how the active plume region expands away from the
  original interface and how salinity, temperature, and vertical velocity can
  approach the relaxation-zone region at different rates.}
  \label{fig:vertical_activity_envelope}
\end{figure}

\begin{figure}[tbp]
  \centering
  \includegraphics[width=\linewidth,height=0.62\textheight,keepaspectratio]{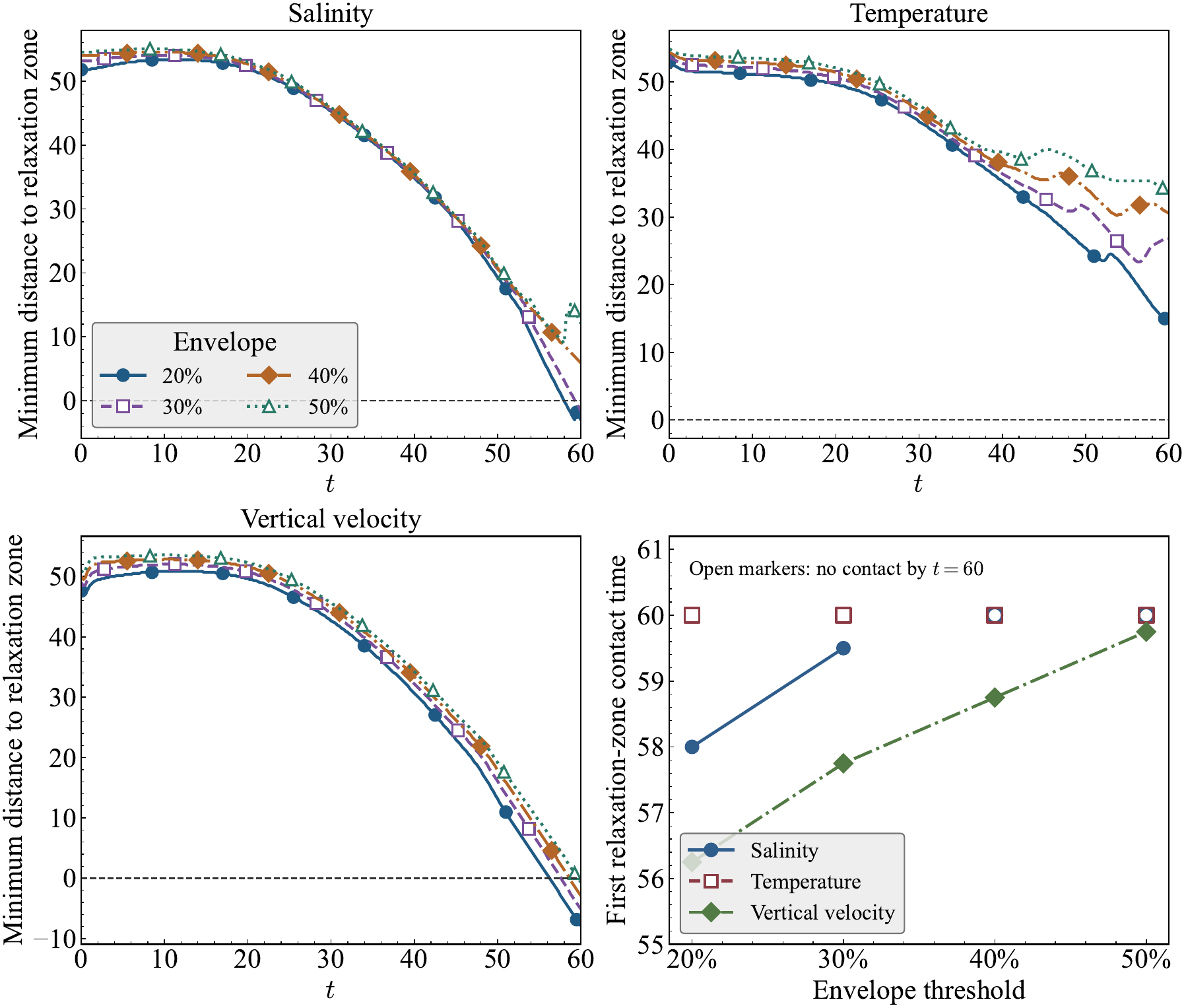}
  \caption{Threshold sensitivity of the mixed-spectrum vertical activity-envelope
  contact measure. Salinity and vertical-velocity contact remain late in the
  simulated window across the tested envelope thresholds, while temperature
  either contacts near the end of the run or remains below contact.}
  \label{fig:vertical_envelope_threshold_sensitivity}
\end{figure}

The vertical activity envelopes shown in
\cref{fig:three_case_vertical_penetration,fig:vertical_activity_envelope,fig:vertical_envelope_threshold_sensitivity},
together with the contact times in \cref{tab:vertical_penetration}, show how the route and interface differences project onto the finite-depth domain. The envelope measure asks where the plume forest is active in \(z\), using a fixed threshold applied consistently across spectra. This matters because a finite-depth interface can either remain a local interfacial mixing event or become a pathway that links the interface to remote layers. \Cref{fig:vertical_penetration_w_width,fig:vertical_penetration_S_width} show the vertical spreading of \(w\) and salinity, while
\cref{fig:vertical_penetration_w_relaxation_distance,fig:vertical_penetration_S_relaxation_distance}
define the boundary-region contact times. The low-mode route reaches the standard relaxation-zone contact threshold first, at \(t=50.5\) for vertical velocity and \(t=51.5\) for salinity. The mixed route approaches later, with first standard contact at \(t=57.75\) for vertical velocity and \(t=59.5\) for salinity. The high-annulus route does not reach the standard contact threshold for either vertical velocity or salinity by \(t=60\).

\begin{table}[tbp]
\centering
\caption{Finite-depth reach measured by the standard activity-envelope threshold. Contact times are the first time the active envelope reaches the inner edge of a relaxation zone; late post-contact values describe boundary approach and are separated from the common interior-comparison window.}
\label{tab:vertical_penetration}
\begin{tabular}{lccc}
\toprule
Case & First \(w\) contact & First \(S\) contact & Final \(w\) width \\
\midrule
Mixed spectrum & 57.75 & 59.5 & 121.6 \\
High-annulus spectrum & n/a & n/a & 96.57 \\
Low-mode spectrum & 50.5 & 51.5 & 152 \\
\bottomrule
\end{tabular}
\end{table}

\Cref{fig:three_case_vertical_penetration} and \cref{tab:vertical_penetration}
show the same ordering in the final envelope widths. The low-mode route connects the interface to remote finite-depth layers earliest, the high-annulus route remains comparatively localized, and mixed forcing has a delayed but real boundary approach. \Cref{fig:vertical_activity_envelope} shows how that mixed-route approach develops, and
\cref{fig:vertical_envelope_threshold_sensitivity} shows that late mixed
contact is not tied to a single cutoff. Interfacial roughness spectrum can therefore control whether salt-finger activity remains near an interface or becomes a pathway for vertical scalar exchange over the available depth. At \(t=60\), the standard vertical-velocity active widths are \(152.04\), \(121.56\), and \(96.57\) for low-mode, mixed, and high-annulus forcing.

The boundary-approach times in \cref{tab:vertical_penetration} also justify the \(t=45\) interior comparison: the ranking is established before the strongest route reaches the standard contact threshold. Later contact times are then a separate finite-depth result.

Late low-mode values are analyzed in this finite-depth context. After \(t=50.5\), the low-mode route is in a boundary-approach regime, so late-time transport and width are not used as interior comparisons.

\subsection{Realization robustness of the mixed-spectrum route}
\label{sec:results_seed_replicate}

\Cref{tab:seed_replicate_robustness} summarizes a second mixed-spectrum
realization that tests whether the velocity-led mixed route is an artifact of one realization. It preserves the same ordering of modal handoff while shifting the exact transition times in both the mid-plane and interior integrated measures. The mid-plane lag changes from \(8.5\) in the baseline realization to \(5.5\) in the replicate, while the interior integrated crossings remain velocity-led in both runs.

\begin{table}[tbp]
\centering
\caption{Mixed-spectrum seed-replicate robustness checks.}
\label{tab:seed_replicate_robustness}
\scriptsize
\setlength{\tabcolsep}{3pt}
\renewcommand{\arraystretch}{1.08}
\begin{tabularx}{\linewidth}{>{\raggedright\arraybackslash}p{0.19\linewidth}>{\raggedright\arraybackslash}X>{\raggedright\arraybackslash}X>{\raggedright\arraybackslash}X}
\toprule
Quantity & Baseline mixed & Seed replicate & Physical implication \\
\midrule
Mid-plane handoff times & w=36.5, S=45, lag=8.5 & w=37, S=42.5, lag=5.5 & Velocity precedes salinity in both realizations. \\
Interior-integrated handoff times & w=39.5, S=47 & w=40.75, S=49 & The handoff also appears in the interior volume. \\
\addlinespace
Salinity dominant wavelength at t=45 & 19.24 & 19.24 (+0.0\%) & The same broad organizing scale is recovered. \\
Salinity effective mode count at t=45 & 86.66 & 98.28 (+13.4\%) & The replicate remains spectrally populated. \\
\addlinespace
Salinity transport at t=45 & 0.07921 & 0.07618 (-3.8\%) & Transport remains same-order and close to baseline. \\
Energy and scalar activity at t=45 & KE=0.1453, w\_rms=0.4951, varS=0.03722 & KE=0.1373 (-5.5\%), w\_rms=0.4824 (-2.6\%), varS=0.03666 (-1.5\%) & All tracked changes stay within the robustness threshold. \\
\addlinespace
First standard relaxation-zone contact & w=57.75, S=59.5 & w=60, S=no contact & The replicate does not become early-contact low-mode-like. \\
\addlinespace
Salinity-gradient interface at t=45 & rms=11.26, thick=5.298, modes=11.78 & rms=11.1 (-1.4\%), thick=5.602 (+5.7\%), modes=12.17 (+3.3\%) & Gradient-based interface geometry remains mixed-like. \\
\bottomrule
\end{tabularx}
\end{table}

\Cref{tab:seed_replicate_robustness} also shows that the replicate preserves
the mixed case's broad scalar organization: the \(t=45\) salinity dominant wavelength is unchanged, the effective mode count remains high, transport and kinetic measures stay same-order, and the salinity-gradient interface metrics remain mixed-like. Interior-comparison salinity transport changes from \(0.07921\) to \(0.07618\), a \(3.8\%\) decrease, while kinetic energy, \(w_{\mathrm{rms}}\), and salinity variance change by \(5.5\%\), \(2.6\%\), and \(1.5\%\).

The mixed-spectrum result shows route robustness with realization-dependent timing. The replicate does not imply that plume patterns or crossing times are deterministic. It does show that the velocity-led route, broad salinity wavelength, strong scalar spectral population, and same-order transport survive an independent phase/noise realization.

The mixed transition times are therefore realization dependent, but the ordering and measured route family are robust.

\subsection{Resolution of the analysis scales}
\label{sec:results_numerical_quality}

The numerical-quality checks in \cref{tab:numerical_quality} require the relevant scales to be separated from the numerical cutoff and advection to remain well within the stable time-step range. The selected-time CFL values are small, high-wavenumber tail fractions are weak, and the dominant planform wavelengths span many grid cells. A higher-resolution mixed-spectrum simulation on a \(512\times256\times1280\) grid gives the same conclusion at smaller grid spacing. Across the three primary spectra, the maximum selected-time CFL is \(0.0515\), the largest horizontal tail fraction above \(0.67k_{\mathrm{Ny}}\) is \(0.00124\), and the minimum selected dominant planform wavelength is \(22.47\) horizontal grid spacings. In the refined mixed-spectrum run, the late-time salinity dominant wavelength is about \(60\Delta x\), while the late-time salinity horizontal tail fraction is \(1.5\times10^{-4}\).

\begin{table}[tbp]
\centering
\caption{Scale-resolution evidence for the three primary spectrum cases and the
higher-resolution mixed-spectrum simulation. Horizontal and vertical tails are
fractions of spectral power above \(0.67\) of the corresponding Nyquist
wavenumber. The higher-resolution simulation assesses numerical resolution of
the mixed-spectrum route measures.}
\label{tab:numerical_quality}
\begin{tabular}{@{}lcccc@{}}
\toprule
Run & Max CFL & Horiz. tail & Vert. tail & Min. \(\lambda_h/\Delta x\) \\
\midrule
Mixed & 0.04888 & \(1.239\times10^{-3}\) & \(1.82\times10^{-8}\) & 22.47 \\
High annulus & 0.04675 & \(1.115\times10^{-3}\) & \(2.051\times10^{-8}\) & 22.47 \\
Low mode & 0.05154 & \(1.047\times10^{-3}\) & \(6.665\times10^{-8}\) & 38.4 \\
Mixed, higher res. & 0.0489 & \(1.50\times10^{-4}\) & \(1.45\times10^{-10}\) & 30.0 \\
\bottomrule
\end{tabular}
\end{table}

\Cref{tab:numerical_quality} shows that the
tracked modal transitions, effective-mode counts, activity envelopes, and interface measures are not being read at the grid cutoff or near an advective stability limit. The refined mixed-spectrum simulation gives the same conclusion at smaller grid spacing. The comparisons therefore describe resolved nonlinear routes within the modeled finite-depth Boussinesq system rather than unresolved grid-scale artifacts.

\section{Discussion}
\label{sec:discussion}

The comparisons show that interfacial roughness is not a passive initial imperfection in finite-depth salt-finger mixing. Its spectrum selects how a thermohaline interface converts displacement into vertical motion, salinity interleaving, scalar transport, and reach toward adjacent layers. Under matched physical parameters, domain, grid, relaxation zones, time step, and analysis measures, short-scale, broad-scale, and mixed roughness produce distinct pathways: branch locking and localization, rapid broad-branch penetration, and delayed velocity-led reorganization with strong scalar spectral broadening.

\subsection{Route selection at a finite-depth interface}
\label{sec:discussion_route_selection}

The finite-depth setting is central to the result. A salt-fingering interface can remain a local interfacial mixing event or become a pathway that connects the original interface to remote layers. Broad initial roughness demonstrates the penetrating outcome, while short annular roughness remains compact and does not reach the standard contact threshold by \(t=60\).

This contrast matters for oceanic and laboratory interfaces whose initial roughness is not spectrally neutral. Wave motion, shear, intrusions, prior mixing, and remnant deformation can leave finite-amplitude structure at selected horizontal scales. The present results suggest that such structure can influence whether fingering remains concentrated near the interface or develops into a vertically penetrating diapycnal exchange pathway. Thus the spectrum changes the visual organization of the fingers and the time over which the interface communicates with the surrounding finite-depth layers.

The short-scale response rules out the expectation that shorter imposed wavelengths necessarily produce a finer and more vigorous finger forest. Under the present finite-depth conditions, the imposed short branch persists, the effective scalar mode count remains small, and the plume forest transports less salt than the broad-route states. The broad-scale response gives the practical scale of the effect: it produces the largest interior-comparison flux and kinetic energy and reaches the finite-depth boundary region first. For an oceanic interface, that distinction is consequential. Two interfaces with the same local density ratio and diffusivity ratio can differ in whether fingering remains a localized interfacial exchange or becomes a rapidly penetrating pathway, simply because the inherited roughness occupies different horizontal scales.

Mixed roughness exposes the mechanism most clearly. It contains short-scale content at the start, yet vertical velocity reaches the broad organizing wavelength before salinity, while salinity fills a much larger population of planform modes than either endpoint. This is why mixed roughness is a distinct route rather than an amplitude interpolation between the endpoint spectra.

\subsection{Branch directionality and plume-passage intermittency}
\label{sec:discussion_directionality_intermittency}

The angular, signed-branch, and asymmetry measures make the route-selection picture more specific. The diagonal or squiggly planform structures are not interchangeable across spectra. The high-annulus and low-mode endpoints are both strongly oriented at the interior-comparison time, and they concentrate power on different branches and at different folded angles. The mixed route has a much lower single-angle anisotropy because competing signed branches coexist. Its visual complexity reflects a different kind of organization, in which broad-route participation, opposite-sign branch content, and a larger scalar spectral population occur together.

Second, local plume-passage asymmetry is distinct from global upper/lower imbalance. In the mixed route, fixed probes and vertical slices can record strongly side-dependent post-transition events. The full-volume interior measures remain much closer to parity, however. The finite-depth forest can therefore have locally intermittent and side-dependent plume passage without becoming a globally one-sided exchange pathway. That distinction is important for the present symmetric finite-depth setup, because it prevents the probe records from being treated as a boundary-imposed or domain-integrated bias.

Together, these measures show that route selection includes branch topology, orientation, scalar-population breadth, local intermittency, volume balance, and radial wavelength transition. The diagonal and locally intermittent structures visible in the fields are therefore part of a measured route, not a separate qualitative observation.

\subsection{Transport strength and scalar spectral richness}
\label{sec:discussion_transport_richness}

The comparison also separates two properties that can be conflated in visual readings of plume forests: transport strength and scalar spectral richness. If the objective is early finite-depth reach and interior-comparison flux, the broad-branch response is strongest in the present set.

Mixed roughness matters for a different reason: it combines broad dominant organization with a far larger salinity effective mode count than either endpoint. It is neither a weak broad-route response nor an average of the two single-band spectra. This separation is important because a broad, high-transport plume field can rapidly connect layers, whereas a scalar-rich field creates a more intricate interleaving geometry and a wider set of local gradients.

This distinction is useful for analyzing salt-finger interfaces in which transport and scalar complexity matter for different reasons. A broad, high-transport plume field can rapidly connect layers, whereas a spectrally rich scalar field can create a more intricate interleaving geometry and a wider set of local gradients. These effects are related and remain separate measures. The gradient-based interface metrics show the separation. Broad forcing gives the largest active-layer displacement and thickening, whereas mixed forcing gives the richest salinity-gradient interface spectrum.

The route-selection claim rests on agreement among complementary measures: modal ratios identify organizing branch, effective mode counts identify spectral population, activity envelopes identify finite-depth reach, and gradient-based interface measures identify the active scalar layer.

\subsection{Active-interface geometry}
\label{sec:discussion_interface_geometry}

The active-interface measures are the bridge between planform route selection and finite-depth exchange. Once fingering has developed, the interface is not represented well by a single scalar zero crossing. The temperature-zero surface gives a direct measure of displaced two-layer geometry, whereas salinity-gradient centroids, thicknesses, and effective mode counts describe the active interleaving layer through which scalar exchange occurs.

Those interface properties do not collapse onto one ranking. Broad forcing produces the largest displacement and thickest active salinity-gradient layer; short annular forcing produces a thinner compact layer; mixed forcing is intermediate in displacement and thickness while producing the richest salinity-gradient interface spectrum. A single zero-crossing height would miss this separation. The gradient-weighted measures capture the active scalar layer after fingering has created interleaving and local salinity-gradient structure.

A route that maximizes interior-comparison flux is therefore not necessarily the route that creates the most populated active scalar interface, and a scalar-rich interface is not automatically the fastest route to remote layers.

\subsection{Finite-depth reach and boundary approach}
\label{sec:discussion_clean_window}

The distinction between interior-comparison dynamics and boundary approach keeps the finite-depth result well defined. The strict three-spectrum comparison at \(t=45\) precedes the first standard low-mode contact with the relaxation zones; later values describe how finite-depth plume forests approach remote layers. The relaxation zones therefore do not set the early route-selection result, but they are part of the later finite-depth problem because the plume forest can eventually approach the far-field regions.

This separation also fixes the role of the vertical relaxation zones. They do not set the early route-selection result, because the main comparison is made before the strongest case reaches the standard contact threshold. They do become part of the later finite-depth problem, because the plume forest can eventually interact with the upper and lower far-field regions. Late-time penetration is therefore a finite-depth boundary approach result, not an unbounded-interface asymptote.

\subsection{Realization robustness of the mixed route}
\label{sec:discussion_seed_robustness}

The mixed-spectrum replicate constrains how strongly the route-selection claim can be stated. It does not reproduce a deterministic plume pattern or identical transition schedule, but it preserves the ordering and character of the route: vertical velocity selects the broad branch before salinity, the dominant salinity wavelength is unchanged at \(t=45\), the effective mode count remains high, and interior-comparison salinity transport changes only slightly.

Changed phases and noise do not turn the mixed spectrum into either single-band response. What survives is the measured route family, not a pointwise plume pattern or deterministic crossing schedule. This is the appropriate robustness statement for a nonlinear plume forest: the exact crossing times shift, while the ordering, broad salinity wavelength, large scalar modal population, same-order transport, and mixed finite-depth reach persist.

\subsection{Resolution evidence and physical scope}
\label{sec:discussion_quality_limits}

The resolution checks place the route-selection comparison away from the numerical cutoff over the analyzed interval. The time-step measure remains small, energetic spectra decay before the highest resolved wavenumbers, and the dominant planform wavelengths are resolved by many grid cells. The route-selection argument depends on spectral movement within the resolved band rather than on energy accumulating at the numerical cutoff. The higher-resolution mixed-spectrum simulation strengthens this point: the same route family is examined on a refined grid whose late-time horizontal and vertical spectral tails remain small and whose dominant planform wavelength remains far from the grid cutoff. These checks support a resolved-scale reading over the analyzed interval, with the conclusions framed around route selection in the modeled finite-depth Boussinesq system.

The physical behavior is consistent with the relevant literature. The imposed density ratio and diffusivity ratio place the interface in the classical fingering regime \citep{stern1960salt,schmitt1994double}. The plume forest evolves past primary columnar structure into a three-dimensional, secondary, intermittent state, consistent with finite-length and secondary-instability expectations \citep{holyer1984stability,kunze1987limits} and nonlinear DNS
\citep{simeonov2009dns}. The scalar transport is measured through
down-gradient salinity flux, following flux-focused fingering studies
\citep{traxler2011dynamics,stellmach2011dynamics}. The interface measures
distinguish plume-forest route selection and active-layer interleaving from a mature staircase sequence \citep{radko2003mechanism,radko2013double}.

The comparison has finite scope. It uses one density ratio, one diffusivity ratio, one primary amplitude family, one finite-depth geometry, and a symmetric far-field relaxation treatment. Within that regime, the comparisons support a precise result. The initial interfacial spectrum selects whether the plume forest remains locked and localized, becomes rapidly broad and penetrating, or follows a delayed velocity-led route with strong scalar spectral broadening. The higher-resolution mixed calculation strengthens the resolved-scale basis of this statement, but the simulations are not presented as a universal grid-converged flux law over parameter space.

\subsection{Implications}
\label{sec:discussion_implications}

The oceanographic implication is that a thermohaline interface can retain dynamically important information about the spectrum of its initial roughness: broad roughness can connect remote layers quickly, short annular roughness can remain compact, and mixed roughness can produce a transport-intermediate but scalar-rich plume forest.

This route-selection view gives a focused way to connect direct calculations, laboratory studies, and observations. Instead of asking only whether salt fingers appear, one can ask which spectral route the interface follows, how quickly the active layer reaches surrounding fluid, and whether scalar structure is concentrated in a few modes or distributed across a broad population. Those questions are directly tied to vertical exchange, finite layer thickness, and the memory of interfacial roughness. They also identify measurable quantities for observations and laboratory comparisons: dominant planform wavelength, active-layer thickness, scalar-gradient spectral population, and the timing of vertical reach.

More broadly, roughness history can matter even when the background thermohaline parameters are fixed. Interfaces in oceans, laboratory tanks, or layered geophysical flows are rarely spectrally neutral; wave strain, remnant shear, intrusive motion, or prior mixing can precondition the nonlinear fingering route. For process studies of oceanic salt fingering, the interface's spectral state is part of the mixing problem. This gives observational and laboratory studies a concrete target: the preexisting roughness spectrum should be treated as a state variable alongside the temperature and salinity jumps when interpreting early finger growth and finite-depth reach.

\section{Conclusions}
\label{sec:conclusions}

The comparisons show that finite-depth salt-finger mixing is not determined only by the thermodynamic instability parameters. Under matched \(\mathrm{Pr}=7\), \(\tau=0.01\), \(\mathrm{R}_\rho=1.2\), domain, grid, amplitude, and relaxation-zone treatment, the imposed interfacial roughness spectrum selects the nonlinear route by which the unstable two-layer interface exchanges salt with the adjacent layers.

High-annulus roughness remains compact and branch-locked through \(t=60\), whereas low-mode roughness selects the broad branch essentially from the beginning. The low-mode route gives the largest interior-comparison salinity transport, the widest interior-comparison vertical-velocity activity envelope, and the earliest standard boundary-region contacts.

Mixed roughness follows a distinct velocity-led route. In the mid-plane measure, vertical velocity reaches the tracked broad branch at \(t=36.5\), while salinity follows near \(t=45.0\). Interior-integrated measures preserve the same ordering, with crossings at \(t=39.5\) and \(t=47.0\). The mixed spectrum therefore follows a route that is not captured by interpolating between the short-annular and low-mode responses.

Transport strength and scalar spectral richness separate. Low-mode forcing is the strongest interior-comparison transport response, while mixed forcing produces the broadest scalar spectral population.

Interface displacement, active-layer thickness, and scalar interleaving are also distinct responses. Low-mode forcing produces the largest active-layer displacement and thickening, whereas mixed forcing produces the richest salinity-gradient interface spectrum.

Finite-depth reach follows the selected route. Low-mode forcing connects the interface to the boundary region first, mixed forcing approaches later, and high-annulus forcing remains below the standard contact threshold through \(t=60\).

Planform directionality and plume-passage asymmetry separate local structure from global balance. Endpoint spectra produce strongly oriented fields on different branches, whereas mixed forcing contains competing signed branches. Local probes and vertical slices record side-dependent plume passage, while full-volume interior measures remain much closer to upper/lower parity.

A second mixed-spectrum realization preserves the route but shifts the precise timing. The mixed route is robust in ordering and measured character, with realization-dependent transition times rather than deterministic crossing times.

The physical implication is that a thermohaline interface can retain dynamically important memory of the spectrum of its initial roughness. Broad roughness can produce rapid finite-depth penetration and strong interior-comparison transport, short annular roughness can remain localized, and mixed spectral content can create a scalar-rich route in which velocity and salinity reorganize on different schedules. Route selection connects interfacial roughness, modal handoff, scalar spectral population, vertical reach, and finite-depth exchange in one physical picture.

Grid and time-step checks support the resolved-scale basis of the simulated interval. High-wavenumber tails remain weak, and dominant planform wavelengths are separated from the grid cutoff. The simulations establish the main physics result: spectral content at a finite-depth two-layer thermohaline interface selects the nonlinear route to salt-finger mixing and therefore changes the timing, geometry, and strength of vertical scalar exchange.

\section*{Data Availability Statement}
The calculations use the open-source Oceananigans framework. A minimal dataset containing mid-plane NetCDF slice outputs and runtime TSV metadata for the four main simulations is archived on Zenodo
\citep{kalathoor2026swfdata}. The complete raw three-dimensional model output
generated for this study exceeds \(5\) TB and is not included in the public archive.

\bibliographystyle{plainnat}
\bibliography{references}

@article{stern1960salt,
  author = {Stern, Melvin E.},
  title = {The salt-fountain and thermohaline convection},
  journal = {Tellus},
  volume = {12},
  number = {2},
  pages = {172--175},
  year = {1960},
  doi = {10.3402/tellusa.v12i2.9378}
}

@article{stommel1956perpetual,
  author = {Stommel, Henry and Arons, Arnold B. and Blanchard, Duncan},
  title = {An oceanographical curiosity: The perpetual salt fountain},
  journal = {Deep-Sea Research},
  volume = {3},
  number = {2},
  pages = {152--153},
  year = {1956},
  doi = {10.1016/0146-6313(56)90095-8}
}

@article{turner1964newcase,
  author = {Turner, John Stewart and Stommel, Henry},
  title = {A new case of convection in the presence of combined vertical salinity and temperature gradients},
  journal = {Proceedings of the National Academy of Sciences of the United States of America},
  volume = {52},
  number = {1},
  pages = {49--53},
  year = {1964},
  doi = {10.1073/pnas.52.1.49}
}

@article{stern1969collective,
  author = {Stern, Melvin E.},
  title = {Collective instability of salt fingers},
  journal = {Journal of Fluid Mechanics},
  volume = {35},
  number = {2},
  pages = {209--218},
  year = {1969},
  doi = {10.1017/S0022112069001066}
}

@article{stern1969layers,
  author = {Stern, Melvin E. and Turner, John Stewart},
  title = {Salt fingers and convecting layers},
  journal = {Deep-Sea Research and Oceanographic Abstracts},
  volume = {16},
  number = {5},
  pages = {497--511},
  year = {1969},
  doi = {10.1016/0011-7471(69)90038-2}
}

@article{turner1974double,
  author = {Turner, John Stewart},
  title = {Double-diffusive phenomena},
  journal = {Annual Review of Fluid Mechanics},
  volume = {6},
  number = {1},
  pages = {37--54},
  year = {1974},
  doi = {10.1146/annurev.fl.06.010174.000345}
}

@book{turner1973buoyancy,
  author = {Turner, John Stewart},
  title = {Buoyancy Effects in Fluids},
  publisher = {Cambridge University Press},
  year = {1973}
}

@article{huppert1981double,
  author = {Huppert, Herbert E. and Turner, John Stewart},
  title = {Double-diffusive convection},
  journal = {Journal of Fluid Mechanics},
  volume = {106},
  pages = {299--329},
  year = {1981},
  doi = {10.1017/S0022112081001614}
}

@article{turner1985multicomponent,
  author = {Turner, John Stewart},
  title = {Multicomponent convection},
  journal = {Annual Review of Fluid Mechanics},
  volume = {17},
  pages = {11--44},
  number = {1},
  year = {1985},
  doi = {10.1146/annurev.fl.17.010185.000303}
}

@book{radko2013double,
  author = {Radko, Timour},
  title = {Double-Diffusive Convection},
  publisher = {Cambridge University Press},
  year = {2013}
}

@article{schmitt1979growth,
  author = {Schmitt, Raymond W.},
  title = {The growth rate of super-critical salt fingers},
  journal = {Deep-Sea Research Part A},
  volume = {26},
  number = {1},
  pages = {23--40},
  year = {1979},
  doi = {10.1016/0198-0149(79)90083-9}
}

@article{schmitt1981form,
  author = {Schmitt, Raymond W.},
  title = {Form of the temperature-salinity relationship in the central water: Evidence for double-diffusive mixing},
  journal = {Journal of Physical Oceanography},
  volume = {11},
  number = {7},
  pages = {1015--1026},
  year = {1981},
  doi = {10.1175/1520-0485(1981)011<1015:FOTTSR>2.0.CO;2}
}

@article{ruddick1983practical,
  author = {Ruddick, Barry R.},
  title = {A practical indicator of the stability of the water column to double-diffusive activity},
  journal = {Deep-Sea Research Part A},
  volume = {30},
  number = {10},
  pages = {1105--1107},
  year = {1983},
  doi = {10.1016/0198-0149(83)90063-8}
}

@article{holyer1984stability,
  author = {Holyer, J. Y.},
  title = {The stability of long, steady, two-dimensional salt fingers},
  journal = {Journal of Fluid Mechanics},
  volume = {147},
  pages = {169--185},
  year = {1984},
  doi = {10.1017/S0022112084002044}
}

@article{kunze1987limits,
  author = {Kunze, Eric},
  title = {Limits on growing, finite-length salt fingers: A Richardson number constraint},
  journal = {Journal of Marine Research},
  volume = {45},
  number = {3},
  pages = {533--556},
  year = {1987},
  doi = {10.1357/002224087788326885}
}

@article{gargett2003differential,
  author = {Gargett, Ann E.},
  title = {Differential diffusion: An oceanographic primer},
  journal = {Progress in Oceanography},
  volume = {56},
  number = {3--4},
  pages = {559--570},
  year = {2003},
  doi = {10.1016/S0079-6611(03)00025-9}
}

@article{you2002turner,
  author = {You, Yuzhu},
  title = {A global ocean climatological atlas of the {Turner} angle: implications for double-diffusion and water-mass structure},
  journal = {Deep-Sea Research Part I},
  volume = {49},
  number = {11},
  pages = {2075--2093},
  year = {2002},
  doi = {10.1016/S0967-0637(02)00099-7}
}

@article{schmitt2005enhanced,
  author = {Schmitt, Raymond W. and Ledwell, James R. and Montgomery, Ellyn T. and Polzin, Kurt L. and Toole, John M.},
  title = {Enhanced Diapycnal Mixing by Salt Fingers in the Thermocline of the Tropical Atlantic},
  journal = {Science},
  volume = {308},
  number = {5722},
  pages = {685--688},
  year = {2005},
  doi = {10.1126/science.1108678}
}

@article{schmitt1994double,
  author = {Schmitt, Raymond W.},
  title = {Double diffusion in oceanography},
  journal = {Annual Review of Fluid Mechanics},
  volume = {26},
  number = {1},
  pages = {255--285},
  year = {1994},
  doi = {10.1146/annurev.fl.26.010194.001351}
}

@article{radko2003mechanism,
  author = {Radko, Timour},
  title = {A mechanism for layer formation in a double-diffusive fluid},
  journal = {Journal of Fluid Mechanics},
  volume = {497},
  pages = {365--380},
  year = {2003},
  doi = {10.1017/S0022112003006785}
}

@article{merryfield2000origin,
  author = {Merryfield, William J.},
  title = {Origin of thermohaline staircases},
  journal = {Journal of Physical Oceanography},
  volume = {30},
  number = {5},
  pages = {1046--1068},
  year = {2000},
  doi = {10.1175/1520-0485(2000)030<1046:OOTS>2.0.CO;2}
}

@article{kelley2003diffusive,
  author = {Kelley, D. E. and Fernando, H. J. S. and Gargett, A. E. and Tanny, J. and Ozsoy, E.},
  title = {The diffusive regime of double-diffusive convection},
  journal = {Progress in Oceanography},
  volume = {56},
  number = {3--4},
  pages = {461--481},
  year = {2003},
  doi = {10.1016/S0079-6611(03)00026-0}
}

@article{radko2014recipes,
  author = {Radko, Timour and Bulters, A. and Flanagan, J. D. and Campin, Jean-Michel},
  title = {Double-diffusive recipes. Part {I}: Large-scale dynamics of thermohaline staircases},
  journal = {Journal of Physical Oceanography},
  volume = {44},
  number = {5},
  pages = {1269--1284},
  year = {2014},
  doi = {10.1175/JPO-D-13-0155.1}
}

@misc{simeonov2009dns,
  author = {Simeonov, Julian A. and Stern, Melvin E. and Radko, Timour},
  title = {Direct Numerical Simulation of 3D Salt Fingers: From Secondary Instability to Chaotic Convection},
  year = {2009},
  eprint = {0910.2434},
  archivePrefix = {arXiv},
  primaryClass = {physics.flu-dyn},
  publisher = {arXiv},
  doi = {10.48550/arXiv.0910.2434},
  url = {https://arxiv.org/abs/0910.2434}
}

@article{traxler2011dynamics,
  author = {Traxler, Adrienne L. and Stellmach, Stephan and Garaud, Pascale and Radko, Timour and Brummell, Nicholas},
  title = {Dynamics of fingering convection. {Part 1}. {Small-scale} fluxes and large-scale instabilities},
  journal = {Journal of Fluid Mechanics},
  volume = {677},
  pages = {530--553},
  year = {2011},
  doi = {10.1017/S002211201100098X}
}

@article{brown2013chemical,
  author = {Brown, Justin M. and Garaud, Pascale and Stellmach, Stephan},
  title = {Chemical transport and spontaneous layer formation in fingering convection in astrophysics},
  journal = {The Astrophysical Journal},
  volume = {768},
  number = {1},
  pages = {34},
  year = {2013},
  doi = {10.1088/0004-637X/768/1/34}
}

@article{garaud2018double,
  author = {Garaud, Pascale},
  title = {Double-diffusive convection at low {Prandtl} number},
  journal = {Annual Review of Fluid Mechanics},
  volume = {50},
  number = {1},
  pages = {275--298},
  year = {2018},
  doi = {10.1146/annurev-fluid-122316-045234}
}

@article{timmermans2008ice,
  author = {Timmermans, Mary-Louise and Toole, John M. and Krishfield, Richard and Winsor, Peter},
  title = {Ice-Tethered Profiler observations of the double-diffusive staircase in the {Canada Basin} thermocline},
  journal = {Journal of Geophysical Research: Oceans},
  volume = {113},
  pages = {C00A02},
  year = {2008},
  doi = {10.1029/2008JC004829}
}

@article{shibley2017arctic,
  author = {Shibley, Nicole C. and Timmermans, Mary-Louise and Carpenter, Jeffrey R. and Toole, John M.},
  title = {Spatial variability of the {Arctic Ocean's} double-diffusive staircase},
  journal = {Journal of Geophysical Research: Oceans},
  volume = {122},
  number = {2},
  pages = {980--994},
  year = {2017},
  doi = {10.1002/2016JC012419}
}

@article{durante2019permanent,
  author = {Durante, Salvatore and Schroeder, Katrin and Mazzei, Lorenzo and Pierini, Stefano and Borghini, Mireno},
  title = {Permanent thermohaline staircases in the {Tyrrhenian Sea}},
  journal = {Geophysical Research Letters},
  volume = {46},
  number = {3},
  pages = {1562--1570},
  year = {2019},
  doi = {10.1029/2018GL081747}
}

@article{stellmach2011dynamics,
  author = {Stellmach, Stephan and Traxler, Adrienne L. and Garaud, Pascale and Brummell, Nicholas and Radko, Timour},
  title = {Dynamics of fingering convection. {Part 2}. {The} formation of thermohaline staircases},
  journal = {Journal of Fluid Mechanics},
  volume = {677},
  pages = {554--571},
  year = {2011},
  doi = {10.1017/jfm.2011.99}
}

@article{ramadhan2020oceananigans,
  author = {Ramadhan, Ali and Wagner, Gregory L. and Hill, Chris and Campin, Jean-Michel and Churavy, Valentin and Souza, Andre and Edelman, Alan and Ferrari, Raffaele and Marshall, John},
  title = {{Oceananigans.jl}: Fast and friendly geophysical fluid dynamics on {GPUs}},
  journal = {Journal of Open Source Software},
  volume = {5},
  number = {53},
  pages = {2018},
  year = {2020},
  doi = {10.21105/joss.02018}
}

@misc{wagner2025oceananigans,
  author = {Wagner, Gregory L. and Silvestri, Simone and Constantinou, Navid C. and Ramadhan, Ali and Campin, Jean-Michel and Hill, Chris and Chor, Tomas and Strong-Wright, Jago and Lee, Xin Kai and Poulin, Francis and Souza, Andre and Burns, Keaton J. and Bishnu, Siddhartha and Marshall, John and Ferrari, Raffaele},
  title = {High-level, high-resolution ocean modeling at all scales with {Oceananigans}},
  year = {2025},
  eprint = {2502.14148},
  archivePrefix = {arXiv},
  primaryClass = {physics.ao-ph},
  publisher = {arXiv},
  doi = {10.48550/arXiv.2502.14148},
  url = {https://arxiv.org/abs/2502.14148}
}

@misc{kalathoor2026swfdata,
  author = {Kalathoor, Sriram P.},
  title = {Data for: Interfacial Roughness Spectra and Finite-Depth Salt-Finger Mixing at a Two-Layer Thermohaline Interface},
  year = {2026},
  publisher = {Zenodo},
  doi = {10.5281/zenodo.20756111},
  url = {https://doi.org/10.5281/zenodo.20756111}
}

\end{document}